\documentclass{egpubl}
\usepackage{eg2021}
 
 \ConferencePaper        % uncomment for (final) Conference Paper
\usepackage[T1]{fontenc}
\usepackage{dfadobe}  

\usepackage{multirow}

\usepackage{cite}  % comment out for biblatex with backend=biber
\BibtexOrBiblatex
\electronicVersion
\PrintedOrElectronic
\ifpdf \usepackage[pdftex]{graphicx} \pdfcompresslevel=9
\else \usepackage[dvips]{graphicx} \fi

\usepackage{egweblnk} 
\usepackage{amsmath}

\usepackage{booktabs} % For formal tables
\usepackage{accents}
\usepackage[normalem]{ulem}
\usepackage[ruled]{algorithm2e} % For algorithms
\usepackage{ulem}
\usepackage{color,soul}
\usepackage{wrapfig, blindtext}
\usepackage{wrapfig}

\SetAlFnt{\small}
\SetAlCapFnt{\small}
\SetAlCapNameFnt{\small}
\SetAlCapHSkip{0pt}

\definecolor{fern}{rgb}{0.2,0.5,0.1}

\definecolor{orange}{rgb}{0.8,0.5,0.0}

\definecolor{fire}{rgb}{0.8,0.2,0.1}

\newcommand{\IGNORE}[1]{}
\graphicspath{{figures/}}

\newcommand{\final}[1]{{#1 }}

\renewcommand{\vec}[1]{\mathbf{#1}} % all vectors are in bold (scalars are in italics)
\renewcommand{\matrix}[1]{\mathrm{#1}} % matrices are upper case non italics
\newcommand{\Anc}{\mathcal{A}} % Ancestor set
\newcommand{\Des}{\mathcal{D}} % Descendant set
\newcommand{\Fun}{\psi} % the function to compute displacement
\newcommand{\RotFun}{\Fun^{R}} 
\newcommand{\TraFun}{\Fun^{T}} 
\newcommand{\TraFunFloppy}{\TraFun_{\text{floppy}}}
\newcommand{\RotFunFloppy}{\RotFun_{\text{floppy}}}
\newcommand{\TraFunSquash}{\TraFun_{\text{squash}}}
\newcommand{\RotFunSquash}{\RotFun_{\text{squash}}}
\newcommand{\PaintFloppy}{k_{\text{floppy}}}
\newcommand{\PaintSquash}{k_{\text{squash}}}
\newcommand{\Vi}{u} % default index of a vertex
\newcommand{\Bi}{i} % default index of a bone
\newcommand{\Bj}{j} % default index of another bone
\newcommand{\Vel}{\vec{v}} % velocity. Note: "speed" is its magnitude, a scalar!
\newcommand{\Pos}{\vec{p}} % position
\newcommand{\RestPos}{\vec{r}} % rest position
\newcommand{\Dis}{\vec{d}} % displacement (i.e. defoemration)
\newcommand{\Mas}{m} % "mass" of a vertex (used only to compute bone centres only)
\newcommand{\AngVel}{\vec{\omega}} % angular vel of a bone
\newcommand{\Wei}{a} % bone-to-vertex link weight
\newcommand{\vWei}{b}
\newcommand{\velweights}{\vWei_i^u}

\newcommand{\AncWei}{\accentset{\leftharpoonup}{\Wei}} % ancestor  version
\newcommand{\DesWei}{\accentset{\rightharpoonup}{\Wei}} % descendant  version
\newcommand{\PosVi}{\Pos^\Vi} 
\newcommand{\VelVi}{\Vel^\Vi} 
\newcommand{\MasVi}{\Mas_\Vi} 
\newcommand{\DisVi}{\Dis^\Vi} 
\newcommand{\RestPosVi}{\RestPos^\Vi} 
\newcommand{\DotPosVi}{\dot{\vec{p}}^\Vi} % time derivative of position
\newcommand{\DotPosViBi}{\dot{\vec{p}}^\Vi_\Bi} % time derivative of position
\newcommand{\VelBi}{\Vel_\Bi} 
 
\newcommand{\PosBi}{\Pos_\Bi} 
\newcommand{\AngVelBi}{\AngVel_\Bi} 
\newcommand{\RotVelBi}{\Vel^R_\Bi} 
\newcommand{\TraVelBi}{\Vel^T_\Bi}

\newcommand{\CentroidBi}{\vec{c}_\Bi}
\newcommand{\PosViBi}{\Pos^\Vi_\Bi} 
\newcommand{\VelViBi}{\Vel^\Vi_\Bi} 
\newcommand{\VelViBj}{\Vel^\Vi_\Bj} 
\newcommand{\WeiViBi}{\Wei_\Bi^\Vi} 
\newcommand{\WeiViBj}{\Wei_\Bj^\Vi} 
\newcommand{\vWeiViBi}{\vWei_\Bi^\Vi}

\newcommand{\AncWeiViBi}{\AncWei_\Bi^\Vi} 
\newcommand{\RotVelViBi}{\Vel_{\Bi}^{R,\Vi}} 
\newcommand{\TraVelViBi}{\Vel_{\Bi}^{T,\Vi}} 
\newcommand{\DesWeiViBi}{\DesWei_\Bi^\Vi} 
\title[Velocity Skinning]%
      {Velocity Skinning for Real-time Stylized Skeletal Animation}

\author[D. Rohmer, M. Tarini, N. Kalyanasundaram, F. Moshfeghifar, M.-P. Cani, V. Zordan]{ {Damien Rohmer$^{1}$, Marco Tarini$^{2}$, Niranjan Kalyanasundaram$^{3}$, Faezeh Moshfeghifar$^{4}$, Marie-Paule Cani$^{1}$, Victor Zordan$^{3}$}\\{$^1$ LIX, Ecole Polytechnique/CNRS, IP Paris, $^2$ University of Milan, $^3$ Clemson University, $^4$ University of Copenhagen}}
\begin{document}

%uncomment for using teaser
\teaser{
\begin{center}
  \includegraphics[width=.93\textwidth]{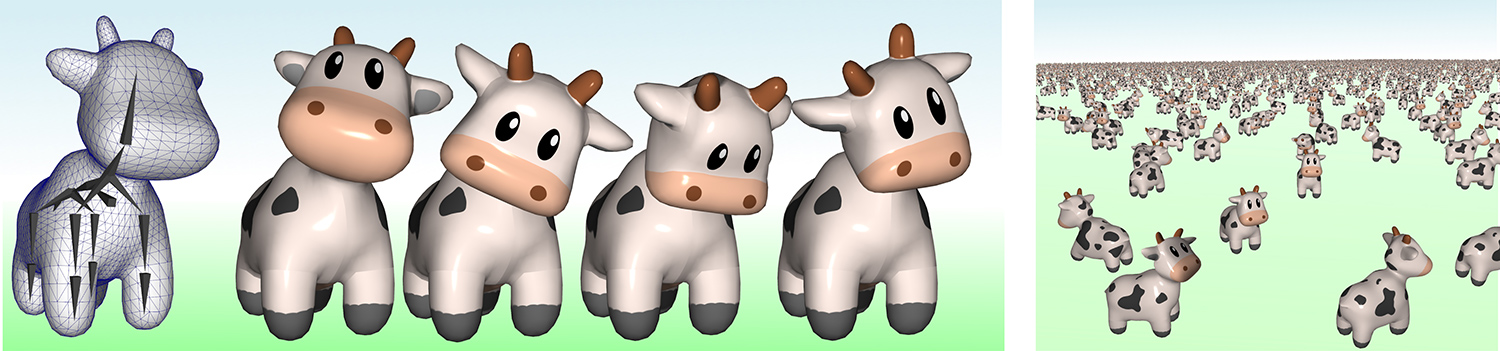}
  \caption{
Left: Skeletal rig, with a single bone in the head: When animated using velocity skinning, secondary animation effects are automatically added to the ear, and face, while the horn can be set as rigid. Right: The native efficiency and simplicity of the method is compatible with GPU implementation used to compute thousands of animated cows in real-time.}
  %Skeletal rig, with a single bone in the head (left).  %all 
  %When animated (right), secondary animation effects are automatically added to the ear, horn, and face, 
  %facial effects are automatically added
  %thanks to the use of velocity skinning.}
  %due to the proposed method.}
  \label{fig:teaser}
\end{center}
}

\maketitle
%-------------------------------------------------------------------------
\begin{abstract} 
%Stylization of skeletal motion usually requires complex rigs.
Secondary animation effects are essential for liveliness.
We propose %a technique that adds
a simple, real-time solution for adding them on top of standard skinning, enabling artist-driven stylization of skeletal motion. Our method takes a standard skeleton animation as input, along with 
%associated 
a skin mesh and rig weights.
%mesh and rig, 
It then derives per-vertex deformations from the different linear and angular velocities along the skeletal hierarchy.  %While the approach is general and customizable, 
We highlight two specific applications of this general framework, namely the cartoon-like ``squashy'' and ``floppy'' effects, achieved from
%can be easily built using 
specific combinations of velocity terms.
%, introducing distinct ``squashy'' and ``floppy'' animation effects. 
%The kinematic approach synthesizes
As our results show, combining these effects enables to
%this method can be used to 
mimic, enhance and stylize
%the appearance of 
physical-looking behaviours within a standard animation pipeline, for arbitrary skinned characters. 
%providing interactive results for arbitrary skinned characters.
Interactive on CPU, our method allows for GPU implementation, yielding real-time performances even on large meshes.
%While interactive rates are achieved on CPU, a GPU implementation allows for large models to be deformed in real-time.  
%We show that it is GPU compatible which 
Animator control is supported through a simple interface toolkit,
%handles %in support of 
%such as weights painted on the 3D character, used to 
enabling to refine the desired type and magnitude of deformation at relevant vertices by simply painting weights. The resulting rigged character 
%adds representative deformations
automatically responds to new skeletal animation, without further input. 
%-------------------------------------------------------------------------
%  ACM CCS 1998
%  (see https://www.acm.org/publications/computing-classification-system/1998)
% \begin{classification} % according to https://www.acm.org/publications/computing-classification-system/1998
% \CCScat{Computer Graphics}{I.3.3}{Picture/Image Generation}{Line and curve generation}
% \end{classification}
%-------------------------------------------------------------------------
%  ACM CCS 2012
 %  (see https://www.acm.org/publications/class-2012)
%The tool at \url{http://dl.acm.org/ccs.cfm} can be used to generate
% CCS codes.
%Example:
%\begin{CCSXML}
%<ccs2012>
%<concept>
%<concept_id>10010147.10010371.10010352.10010381</concept_id>
%<concept_desc>Computing methodologies~Collision detection</concept_desc>
%<concept_significance>300</concept_significance>
%</concept>
%<concept>
%<concept_id>10010583.10010588.10010559</concept_id>
%<concept_desc>Hardware~Sensors and actuators</concept_desc>
%<concept_significance>300</concept_significance>
%</concept>
%<concept>
%<concept_id>10010583.10010584.10010587</concept_id>
%<concept_desc>Hardware~PCB design and layout</concept_desc>
%<concept_significance>100</concept_significance>
%</concept>
%</ccs2012>
%\end{CCSXML}

%\ccsdesc[300]{Computing methodologies~Collision detection}
%\ccsdesc[300]{Hardware~Sensors and actuators}
%\ccsdesc[100]{Hardware~PCB design and layout}

\printccsdesc   
\end{abstract}  
%-------------------------------------------------------------------------
\section{Introduction}

Trained animators bring 2D characters to life 
through the inspired
%careful 
application of a few well-known principles, such as \emph{squash, stretch} and \emph{follow through}~\cite{illusionOfLife,Lasseter,AnimatorHandbook}. The latter are used to enhance the physicality of motion and make cartoon figures more expressive, with emotion and appeal. Unfortunately, the application of these rules
%of these rules 
to 3D~animation is quite challenging.
%synthesis can be problematic, 
Indeed, the stylization and subtlety of the desired motion is at odds with the assumptions of existing pipelines~\cite{ves_handbook}, such as linear blend or dual quaternion skinning techniques (LBS and DQS, respectively), where skin deformation is directly derived from the current skeletal pose.
Further, the rules themselves 
%are anecdotal and 
require interpretation to be extended to 3D, which makes them difficult to formalize into generic algorithms~\cite{Yong_TVCG_2012, Robert_ICIVC_2013, Dvoroznak_SIGGRAPH_2017} (among others, see Section~\ref{sec:related-works}).

%The focus of this paper is a technique for adding controllable tools to existing pipelines to support the automatic synthesis of deformations that animators can manipulate to exhibit classic effects of animation. 
We introduce a novel method to enhance 3D character animation pipelines with secondary motion control. Our method is easy to customize, handles the main animation principles, and supports fine-grain animator control.
%In practice, the most prevalent approach to applying deformation to characters is through skinning, for example linear blend skinning (LBS) or dual quaternion skinning (DQS) in which skeletal joint animation drives the blended skinning surface that is rendered. 

While LBS and DQS derive skin deformation from the current, static configuration of skeletal joints, we claim that considering the additional influences of translational and rotational velocities along the skeletal structure is essential to bring more diverse and lively skin responses.
%We propose an extension to basic LBS which incorporates a set of additional influences on the blended vertex, corresponding to translational and rotational effects derived from velocity.
{\em Velocity skinning} consists in adding this set of additional influences to a standard skinning method (LBS, in our implementation). This insight enables us to synthesize dynamic-looking, controllable skin deformations directly from the current state of skeletal motion. In particular, it enables us to achieve \emph{squash \& stretch} and \emph{drag}-like exaggeration for \emph{follow through}, among others (note that for sake of simplicity, we reuse these well-known terms from animation principles~\cite{illusionOfLife}, to denote the similar effects generated by our system). 
%Note, \emph{squash \& stretch}, \emph{follow through}, and \emph{drag} are key terms defined in the existing literature, e.g. ~\cite{illusionOfLife}, and thus we adopt them to describe similar effects generated by our system. 

Figure~\ref{fig:teaser} illustrates this framework with a single animated bone, showing that it can bring various dynamic effects, such as the floppy ears and squashy face of the cow. While skilled animators could use key-frames to manually add such secondary motion, velocity skinning directly encodes them within the skinning pipeline. Thanks to GPU implementation, computed as a single-pass vertex shader, velocity skinning can be applied in real-time to any skeleton-based animated character, independently from its animation and in a reusable manner.
%, akin to the benefits derived from skeletal LBS.

Our method uses, as input, a standard skinned mesh with a skeleton animation. It generates a deformation of the mesh, expressed as per-vertex displacements, that emphasizes (and stylizes) motion induced by the movement of the skeleton. We take advantage of the hierarchical nature of the skeleton to approximate dynamic deformations that otherwise, would not be accessible without either some manual input from a skilled animator or an expensive, physically-based simulation.
%obtained short of direct animation by a skilled animator or a comprehensive, and likely expensive, physical simulation.
%MP ??????
The key idea is that the skeleton hierarchy embedded in the skinning computation provides enough information to decompose the motion of the mesh into meaningful, easily directable sub-motions that can be automatically associated with deformations.
We contrast this approach to existing modeling tools, such as the deformers in a standard animation software package, which require manual setup and keyframing by a skilled artist for every animation, which is both time consuming and cumbersome.

Although the general velocity skinning framework supports a wide set of deformations, in this paper we showcase the utility of two specific characteristic deformations that we term ``squashiness'' and ``floppiness''. The magnitude of these two deformations effects can be finely tuned with scalar weighting parameters to reflect the different properties of the physical materials
%.Following the paradigm to control standard blending in smooth skinning, weight influence for each 
% Damien: no this is simpler than skinning weights: 1 value per vertex only, no barycentric problem here.
, and can be defined as per-vertex attributes ``painted'' on the mesh by an artist. Along with a small toolkit of interface handles, this allows simple control over an intuitive space of possible deformation behaviors, for example, to produce effects that appear
to be made from heterogeneous materials.

The key contributions of this work are the following: we add a new term to the standard skinning formulation that creates customizeable spatial deformations based on velocity; we present a weighting scheme that derives consistent skeletal weights for the proposed framework from traditional (LBS) skinning weights; we split the translational and rotational velocity influences and show how they can be combined to create different effects, e.g. \emph{squash \& stretch} and (floppy) \emph{follow through / drag}; and finally, we develop a set of deformation handles in support of making the effects controllable, based both on the desired output and the distinct characteristics of a given rigged skeleton.

We tested velocity skinning on both CPU and GPU implementations, to show its scalability to large meshes. A reference real-time web application is openly available at \textbf{ \href{https://velocityskinning.com}{https://velocityskinning.com}}, and remains anonymous for reviewing purpose. 
%We also note that 
The supplementary material associated with this submission contains the source code of the web application as well as the entire C++ and GPU interactive interface.

\section{Related Work} \label{sec:related-works}

%\Damien{Damien: The related work has been globally rewritten}

% Generic skinning methods (our method can possibly be applied to then)
Skin deformation of animated characters can be seen as  %largely 
mostly rigid, being driven by bones. 
Linear blend skinning (LBS) offers a simple solution for capturing this behaviour while enabling to smoothly blend deformations 
%between regions driven by different bones, thanks to its ability to blend deformations 
near joints~\cite{magnenat1988joint}. 
%. The technique offers smooth transitions between bones by blending a small region near the bone joints~\cite{magnenat1988joint}. 
Being fast and highly customizable, LBS is routinely used in the animation pipeline. Dual quaternion skinning~\cite{kavan2008geometric} is another popular choice, which solves some of the artifacts arising from 
%rotation blending
LBS in case of high rotation angles, thanks to a non-linear blending mechanism that can be
%non linear transformation blending can be
computed efficiently on the GPU. 
Encoding 
%and retrieving 
surface details with respect to a simpler deformed surface 
%were 
was also exploited to ease rig generation while improving visual results in general. For instance Delta Mush~\cite{mancewicz_2014_digipro_delta_mush} uses Laplacian smoothing to generate such simple surface and encodes the details in a local reference frame to compute visually pleasing skinning very efficiently~\cite{huy_2019_siggraph_ddm}. 
%On the other side, %approximating the surface as a constant isovalue of a scalar field
%%% MP: Untrue! Implicit skinning stores a different field value at each vertex, enabling to re-project each of them on a different isosurface and retrieve details.
In contrast, implicit skinning ~\cite{vaillant_2013_siggraph_implicit_skinning} approximates skin as the iso-surface of an implicit field (the blend of fields associated with different bones) in which the mesh is embedded, each vertex storing its own iso-value to preserve details. The use of advanced blending operators enables not only to preserve volume at joints, but also to avoid inter-penetrations and achieve bulging skin in contact regions. This framework was extended to account for skin sliding effects~\cite{vaillant:Siggraph-Asia-2014}. \final{Energy-based formulations associated with a set of positional, and possibly rotational constraints, were used to infer~\cite{jacobson2012} and increase the range of possible deformations~\cite{wang2015}.}
%Moving away from standard LBS, energy-based formulations associated with a set of positional, and possibly rotational constraints, were used to generate more general, artifact-free deformations~\cite{jacobson2012, wang2015}. 
All these approaches 
%allow for general improvement of skinning deformation 
do improve skin deformation in their own ways, but are not able to imitate dynamic behaviour, as our new method does. They could be combined with our work by serving as better input for the velocity-based deformer.
%and could be possibly serve as better input for the velocity-skinning method,
%our velocity-based deformer, 
%although we will 
In this paper, we rather derive velocity-based deformations
%the core of our relation 
from the standard LBS formulation, in order to remain fully compatible with standard production pipelines. 

% Skinning improvement that requires to modify the model (possibly not pipe-line compatible)
Skinning has also been subject to other improvements, related to an increase of the number of degrees of freedom, in order to fulfill specific criteria~\cite{jacobson_star_2014}. To mention only a few, this family of methods includes the interpolation in pose space~\cite{Lewis_siggraph_2000}, the automatic insertion of extra bones~\cite{mohr2003}, the addition of extra skinning weights~\cite{wang2002multi,merry2006animation}, the integration of limbs scaling~\cite{jacobson2011stretchable}, or local swing and twist deformers extracted from the blended bone transformations~\cite{kavan_siggrapha_2012}.
Allowing to deform the rest poses can also improve skinning. Such deformation can be automatically computed from a principal component analysis applied to a target model, such as a detailed finite element simulation for instance~\cite{Kry_SCA_2002}.
Curved skeletons~\cite{yang2006curve} were used to provide smoother skinning results and solve artifacts,
%continuous representation of skinning, 
%solving its common artifacts, 
and has been combined with extra deformers~\cite{forstmann2007sca}. \\
These method however depart from the standard production pipeline. 
%One should however note that standard production pipeline
The latter~\cite{ves_handbook} is subject to strict constraints including computational efficiency as well as not easily accommodating changes affecting weights or the skeletal structure. Skeletons for which complex rigs may be scripted~\cite{nieto_siggraph_talk_2016}, and possibly shared through multiple characters, should typically not be modified, as they are the core of animation assets. In addition, constant scalar skinning weights painted by skilled artists are the de-facto standard of production-compatible skinning methods. Unfortunately, this pipeline maintains the animated skin as a purely passive geometry element, merely following the skeleton with no dynamic deformation effects.

% Comparison to standard geometric deformers in current tools
In contrast, the principles of animation~\cite{illusionOfLife}, based on exaggeration and stylization of physical phenomena, have helped artists bring their characters to life for many decades. Since such ``cartoon physics'' is not supported by standard motion pipelines, professional animators often rely on additional deformers~\cite{maya_deformer} -- for "squash", "jiggle", "bend", and so on. These act as extra, customized layers, designed to produce a wide variety of effects, but need to be tuned by adjusting their influence over mesh vertices from manual setting or procedural functions to be defined, as well as setting and attaching their possibly varying magnitude over the keyframes along the animation. Our method can be seen as an automatic parameterization of such deformers for cartoon like effects. It is readily available on rigged-animated model as it seamlessly makes use of existing skinning weights to set a model-aware influence along the mesh geometry, and rely on skeleton velocity to automatically adapt the magnitude of the deformation over time or keyframes. 
In addition, our approach still remain compatible with fine-grain artistic control from a per-bone control to magnitude weights painted at the per-vertex level if needed.

% Input-based cartoon effect
Computer graphic researchers have offered a variety of approaches related to ``cartoon physics''. Largely, the approaches allow the cartoon-like effects to be applied to the shape using additional inputs, for example, natural extensions of 2D effects can be achieved through the use of sketch-based interfaces guiding computer-generated deformation such as exaggeration~\cite{Li_SCA_2003}, geometric constraints~\cite{nealen_siggraph_2005,rohmer_sca_2009}, up to guiding an entire suggestive animation~\cite{Kazi_UIST_2014,Kazi_CHI_2016}. Example-based techniques have also been explored to model arbitrary predefined deformations~\cite{Robert_ICIVC_2013,Dvoroznak_SIGGRAPH_2017,Ruhland_CS_2017} or rendering styles~\cite{Benard_SIGGRAPH_2013} that can be triggered during animation and transferred to a target shape~\cite{Bregler_2002, Young_GM_2012}. These approaches provide a fine level of control and artistic expressiveness on the visual result, but they must be set up manually for each specific shape and animation. 
Regression based approaches were also used to generate secondary effects from simulations examples~\cite{aguiar2010}, or captured humans motions~\cite{pons-moll2015}, but were not used for exaggerated deformation.

% Full physically-based models
%Dynamic-based deformations, 
Physically-based deformations, on the other hand, naturally handle automatic %time-related 
dynamic behaviours. Time integration as well as elastic energy formulation is widely used to improve skin deformation~\cite{deul_vriphys_2013} as well as for avoiding self collision~\cite{capell_sca_2005, mcadams_siggraph_2011}, and can be triggered by skeletal animation~\cite{capell_siggraph_2002}. 
Efficient computation can be obtained from the use of subspaces modeling adapted deformation modes~\cite{james_tog_2002} and computed for instance in the rig space~\cite{hahn_sca_2013,RedMax2019}, using helper bone controller~\cite{mukai2016}, or directly in pose space~\cite{xu_siggraph_2016}. Position and projective based dynamics~\cite{muller2005, bouaziz2014, macklin2016} are also popular physically-inspired methods, allowing real time deformation for moderately detailed shapes while being able to handle arbitrary non-linear constraints. They can successfully be used to animate dynamic characters~\cite{Rumman_PPD_skinning_2014, komaritzan_i3d_2018} and even incorporate efficient voxels-based layers of bones, muscle and soft tissues~\cite{iwamoto2015}.
Interestingly, custom physically-based models were also developed to exaggerate specific cartoon-like effects~\cite{Garcia_VRIPHYS_2007, Coros_SIGGRAPH_2012, Bai_SIGGRAPH_2016} on arbitrarily animated models.
%, as well as adding elasto-dynamic secondary effects in the orthogonal subspace of the rig displacement~\cite{zhang2020}. 
\final{Zhang~et al.~\cite{zhang2020} propose to enrich skinning animation with dynamic secondary effects. Their technique generates oscillation, follow through, and even collisions, that complement primary animations by computing these in the orthogonal subspace of the rig. In contrast, our system focuses on simpler effects aimed at an artist-driven workflow.  While their technique requires a few seconds per frame for a mesh of a thousand elements, ours is interactive with much bigger models to support an artist's needs.  We further note that velocity skinning does not aim solely at secondary effects as expressed by physical laws, rather the approach extends and exaggerates the motion in a flexible, general manner, especially integrating artist-directable parameters compatible with cartoon-like animation principles.}

%DAMIEN: Recently,Zhang  et  al.  [ZBLJ20]  proposed  a  physics-inspired  approach  toenrich  skinning  animation  with  secondary  effects.  It  brings  fleshoscillations and follow through effects that complements the mainartist directed motion in computing the added dynamic effects inthe orthogonal subspace of the rig. This approach allows very richdynamic deformation well suited to animate shape details possiblyhandling contacts, but requires a few seconds of computations perframe for a mesh containing a thousand of elements. Our method,on the contrary, provides simpler shape deformation, but is an or-der  of  magnitude  faster  thanks  to  the  use  of  closed  form  proce-dural deformers, allowing real time animation up to a million ofelement. We further note that velocity skinning doesn’t primarilyaim at proposing full secondary effects added with rich animateddetails in addition to a main motion as these effects are naturallyexpressed by physics-laws, but rather extends and exaggerate themain motion itself in integrating artist-directable parameters com-patible with cartoon-like animation principles.

% Simple physical "layer" on top of the model
%\sout{As full physically-based simulation can remain difficult to author and often too costly for animation pipelines~\cite{huy_2019_siggraph_ddm},} 
Simpler dynamic deformers have also been explored, from early work coupling particles and implicit surfaces to achieve {cartoon-physics} effects~\cite{Opalach_WorkshopAnimSimu_1994}, through extra bones attached to the skeleton to model flesh oscillations~\cite{larboulette_2005}, or %splitting skeleton bones into 
the use of sub-bones connected by springs to achieve curvy, dynamic shapes~\cite{Kwon_CGF_2008}. Muscles approximations have also been developed both in explicit~\cite{ramos_2013} and implicit~\cite{rousselet_2018} formulations. Similar to our approach, kinodynamic skinning~\cite{angelidis_sca_2007} proposes an integrated velocity-based formulation for skinning that can be tuned to exaggerate dynamic visual effects. While their deformation is expressed as a vector field to ensure fold-over-free trajectories for vertices, this requires costly numerical integration along streamlines which makes it computationally prohibitive.

% Kinematics approaches
In contrast, our method belongs to kinematic approaches, i.e. those that use position trajectory, or directly velocity and acceleration information, without requiring any time integration. In the specific case of predefined motion, \final{time-based filter applied to the vertex position trajectory~\shortcite{Wang_SIGGRAPH_2006}, or time-wrapper applied to the bone motion~\shortcite{Kim_CGI_2006}, were proposed to}
%Wang et al.~\shortcite{Wang_SIGGRAPH_2006} proposes a time-based filter applied to the vertex position trajectory able to 
express exaggerated motion in space, as well as the notion of \emph{follow through} and \emph{anticipation} over time. The use of oscillating splines~\cite{kass2008} aim at deforming the existing animation curves to mimic the trajectory of a damped mass-spring model. Associated to a phase shift along the surface geometry, the method can model drag effect followed by wiggling motion. The magnitude and phase field should however be parameterized by the user on every given model and do not automatically take advantage of the existing rig. \final{Note that vertex trajectory filtering-based approaches can be seen as complementary of our method in caputing effects such as follow through effect that could be combined with ours.}\\
Geometrically defined \emph{squash \& stretch} effects triggered by kinematics and collisions were also tackled by deforming the bones themselves~\cite{Yong_TVCG_2012}, as well as slowing down trailing joints. %The squash effect was however applied to the entire character in a global way, enabling little control. 
\final{While their approach allows local elongations along the bones, they only support global scaling with respect to the root-bone velocity. Therefore the space of deformations is more restricted than in our work which allows scaling and bending limbs along arbitrary directions computed from bone velocities.}
Closer to our work, Nobel et al.~\shortcite{Noble_GRAPHITE_2006} developed a specific bending deformer able to curve limbs based on the direction of motion. %Similar to our method, bone velocity and some geometric criteria are used as deformer parameters, but the method only applies to a single bone for each deformation, while our approach is able to smoothly apply a deformation across several bones, thanks to the use of a skinning-like formulation.
\final{Similar to our method, bone velocity and geometric criterion are used as deformer parameters to bend limbs. However, the deformation for a given vertex is based solely on the attached bone, without considering the global hierarchy. Their velocity and deformation parameters are only computed with respect to a given bone and its parent, which lead to artifacts between joints. In contrast, our approach produces seamless deformations that remain coherent for arbitrary skeletal hierarchies.}

% Conclusion
Our method relies on the use of bone velocities to define closed-form, generic geometric deformers. As such, and similarly to static deformers, neither subsequent motion nor past configurations are required, which makes the method easier to add to a standard pipeline, at low cost, and greatly eases tuning.
In addition, our 
%geometric 
deformations do not require any change to the input animation skeletons, while still being able to synthesize curvy skin shapes as well as 
%the appearance of 
mimicking time-delayed deformations. These smooth and coherent deformations are achieved thanks to a new formulation built on standard skeletons and skinning weights, as presented next.
\section{Velocity skinning}

We introduce our velocity skinning formulation by drawing from linear blend skinning (LBS). 
%However, it can also be  
It could alternatively be
directly plugged into another form of skinning such as DQS, as discussed
%as we explain in the implementation section (see 
in Section~\ref{sec:implementation}. 

In LBS, at each frame, the position $\PosVi$ of vertex $\Vi$ is computed as a weighted sum of the transforms associated to each bone applied to the rest pose $\RestPosVi$ vertex as
\begin{equation}
\label{eq:skinning}
\PosVi = \left( \sum_\Bi \WeiViBi \, \matrix{T}_\Bi \right) \RestPosVi \,,
\end{equation}
where $\matrix{T}_\Bi$ is the current frame's transform for bone $\Bi$, obtained by accumulating rotations and translations along the hierarchical skeleton structure. Equivalently, $\PosVi$ can also be understood as a linear combination of positions $\PosViBi = \matrix{T}_\Bi\, \RestPosVi$ of the vertex $\Vi$ moved according to bone $\Bi$ as
\begin{equation}
\label{eq:skinning2}
\PosVi = \sum_\Bi \WeiViBi \PosViBi.
\end{equation}
As the skeleton is animated by changing rotations (and sometimes translations) of bone joints at each frame, LBS moves vertex $\Vi$ to trace a trajectory over time. 

For velocity skinning, we base the foundation of our approach on the premise that, like vertex position, vertex velocity $\VelVi = \DotPosVi$
can also be understood as a linear combination of a set of \emph{component velocities} $\VelViBi$.
We exploit this by proposing to make component velocities induce separate, individually weighted displacements to the vertex, while allowing each velocity's influence to be customizable (to add distinct animation effects) through a function, $\Fun$. The velocity-based position displacements are then added through linear combination to the final procedural mesh deformation, as displacement $\DisVi $. That is,
\begin{equation}
\label{eq:displacements}
\DisVi = \sum_\Bi{ \velweights  \Fun_{\Bi} ( \VelViBi ) } \, ,
\end{equation} 
where $\velweights$ are the bone weights defined per vertex. Deformer functions $\Fun_\Bi$ take in account the geometry of bone $\Bi$.

The LBS position $\PosVi$ from Equation~(\ref{eq:skinning}), which is determined by the \emph{static} pose, is displaced by $\DisVi$ from Equation~(\ref{eq:displacements}), obtaining an additional mesh deformation which is automatically induced by the skeletal \emph{animation}. 
\begin{equation}
\PosVi{}^\prime = \PosVi + \DisVi \,.
\end{equation}

\noindent
To take advantage of the existing skinning weights from LBS, we show how we derive $\velweights$ from the existing skinning weights $\WeiViBi$
in Section~\ref{sec:decompose}. To support the development of the desired velocity effects, we show in Section~\ref{sec:vel} a breakdown of velocity that separates the influences of translational and rotational elements. Finally, in Section~\ref{sec:deform}, we illustrate a general framework for the function $\Fun$, which can be employed to achieve different animation effects as well as combine the inputs of multiple effect \emph{deformers}.

\subsection{Velocity component weighting} 
\label{sec:decompose}
While a general formulation for velocity skinning could introduce an arbitrary weighting scheme for $ \vWeiViBi $ in combining the velocity influences in Equation~\eqref{eq:displacements}, instead, we opt to take advantage of LBS by reusing the skinning weights already attached to the original mesh.  To this end, we derive the velocity weights directly from the LBS weights and employ them in computing displacements.

%\Damien{(I rewrote this paragraph that was confusing for reviewer1) 
Let us first consider the 
%rigidly 
deformed position $\PosViBi$ 
%as 
defined by Equation~\eqref{eq:skinning2}, and decompose the velocity $\DotPosViBi$ of vertices along the skeleton hierarchy. We call vector $\VelViBi$ the \emph{component velocity} induced by the rigid motion of bone $\Bj$ relatively to its immediate parent. A straightforward decomposition along this kinematic chain shows that $\DotPosViBi$ can be expressed as the sum over all these relative velocities such that
\begin{equation}
\label{eq1}
 \DotPosViBi = \sum_{\Bj \in \Anc(\Bi)}{ \VelViBj } \,,
\end{equation}
%\Damien{
where $\Anc(\Bi)$ denote the set of all ancestors of bone $\Bi$ (including $\Bi$ itself). Note that all velocity vectors are expressed here in the same global reference frame.
%}
%\sout{Assuming rigid body skinning, a simple decomposition of the kinematic chain shows that the velocity $\VelVi$ of vertex $\Vi$ is given by the rigid motion of the associated bone $\Bi$ with respect to its parent $\Bj$, combined with the rigid motion of bone $\Bj$ with respect to its parent, and so on until the root bone is reached. Thus, if we denote the set of ancestors of bone $\Bi$ (including $\Bi$ itself) as $\Anc(\Bi)$, we can compute the linear velocity $\VelVi$ of vertex $\Vi$ as }
%\begin{equation}
%\label{eq1}
 %\VelVi = \sum_{\Bj \in \Anc(\Bi)}{ \VelViBj }.
%\end{equation}
%\sout{with all velocities being expressed in the  global reference frame.}

\begin{figure}
\resizebox{0.85\linewidth}{!}{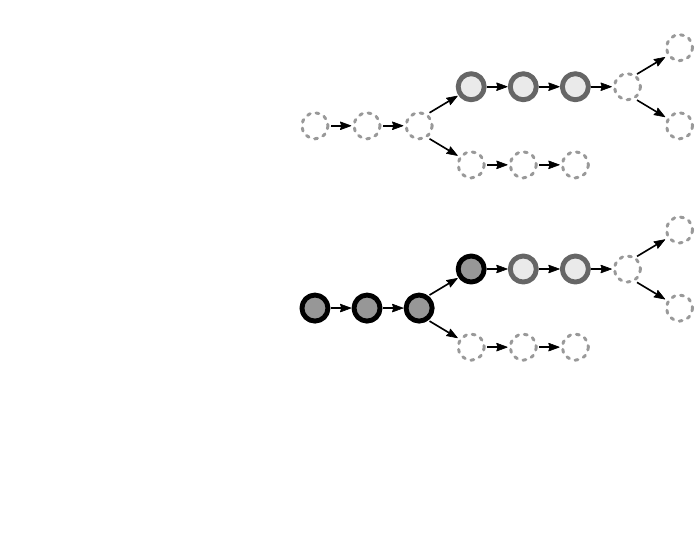}
\caption{ An example of the upward-propagated  weights (middle) and downward-propagated weights (bottom) derived for the original bone-to-vertex weights (top), for a given vertex $\Vi$, as for Equations~(\ref{eq:AncWei}) and (\ref{eqBarycenters}), respectively. For smooth skinning, each vertex is normally linked to a short sequence of interconnected bones (top). In velocity skinning, we use upward-propagated weights (middle) that can be derived from the original weights.  Note, most ancestors have an influence of~1 on the given vertex. Bones $\Bi$ are represented as circles, and arrows represent the hierarchical structure of the skeleton.}
 \vspace{-2em}
\label{fig:propweights}
\end{figure}

%In LBS, multiple bones may influence the same vertex $u$ through weights for each bone $\Bi$, with constant weights $\WeiViBi$ summing up to 1 for each vertex (See Figure~\ref{fig:propweights}, top).(\Damien{Maybe put this discussion of sum to 1 later, when we discuss the "upward-propagated weight"}) 
%\sout{Applying Equation~\eqref{eq1} to each linked bone, and weighting the associated velocity with the respective weight, we get}
Differentiating  Equation~\eqref{eq:skinning2} with respect to time, and plugging Equation~\eqref{eq1} into it leads to
\begin{equation}
\label{eqVelSkinning}
 \VelVi = \sum_\Bi{ \WeiViBi \left( \sum_{\Bj\in \Anc(\Bi)}{ \VelViBj } \right) }.
\end{equation}
\noindent This can be rewritten as a single summation (see Appendix~\ref{sec:derive} for the full derivation):
\begin{equation}
\label{eqVelSkinning2}
  \VelVi =  \sum_{\Bi}{ \AncWeiViBi \VelViBi } 
\end{equation}
where the \emph{upward-propagated} weight vector $\AncWeiViBi$ (see Figure~\ref{fig:propweights}, middle) is defined at each vertex $\Vi$ as
\begin{equation}
\label{eq:AncWei}
 \AncWeiViBi =  \sum_{ \Bj \in \Des(\Bi)}{ \WeiViBj }
\end{equation}
\noindent with $\Des(\Bi)$ denoting the set of descendants of bone $\Bi$, i.e. the set of bones in the subtree rooted in bone $\Bi$ (including $\Bi$).
Equation~(\ref{eqVelSkinning2}) is the velocity counterpart of Equation~(\ref{eq:skinning2}). To create consistency in the skinning weights of LBS and velocity skinning we assign $\velweights$ to $\AncWeiViBi$ in Equation~(\ref{eq:displacements}). 
Note that in LBS, \(\sum_{\Bj}\WeiViBi=1\) for every vertex \(u\), while this is not the case for the \emph{upward-propagated} weights. Still each individual weight \(\AncWeiViBi\) belongs to the interval \([0,1]\).
%multiple bones may influence the same vertex $u$ through weights for each bone $\Bi$, with constant weights $\WeiViBi$ summing up to 1 for each vertex (See Figure~\ref{fig:propweights}, top).

\subsection{Velocity component estimation} 
\label{sec:vel}

To support customization in the procedural use of the velocity deformers, we decompose velocity into its translational and rotational components.
That is, for every vertex $\Vi$, the velocity component associated to each bone $\Bi$ is given by 
\begin{equation}
\label{eq:velDecompose}
\VelViBi = \RotVelViBi + \TraVelViBi.
\end{equation}
%where $\RotVelBi$ is the linear velocity induced by the \emph{rotation} motion of the joint of bone $\Bi$, and $\TraVelBi$ is the linear velocity induced by its \emph{translation} motion (when present).
\final{where $\RotVelBi$ is the vertex velocity component induced by the \emph{angular velocity} of the joint of bone $\Bi$, and $\TraVelBi$ is the vertex velocity component induced by the bone's \emph{linear velocity} (when present).}

To evaluate the terms of Equation~(\ref{eq:velDecompose}), we must extract the angular $\AngVelBi$ and linear velocity $\VelBi$ of each bone $\Bi$. First, we extract them in the local space of the parent bone, then we propagate them upward in the skeleton hierarchy (using forward kinematics) to express them in the global reference frame.
The actual computation of these velocities can either be carried through analytic derivation when 
%the trajectory 
the animation is provided as 
a set of parametric 
% animation 
curves, or via finite differences between previous and current frames to easily support interactive deformation along \emph{mouse motions}.
%CENTERED VERSION

\hspace{-4mm} 
%MP: keep this to suppress the indentation
\begin{minipage}{0.5\linewidth}
Vector $\RotVelViBi$ is then  the  linear velocity for vertex $\Vi$  in position $\PosVi$ induced by the angular rotation $\AngVelBi$ around bone origin $\PosBi$:
\begin{equation}
 \RotVelViBi = \AngVelBi \times ( \PosVi - \PosBi) \,.
 \label{eq:joint}
\end{equation}
\end{minipage}
\begin{minipage}{0.45\linewidth}
%% Creator: Inkscape inkscape 0.92.4, www.inkscape.org
%% PDF/EPS/PS + LaTeX output extension by Johan Engelen, 2010
%% Accompanies image file 'schema.pdf' (pdf, eps, ps)
%%
%% To include the image in your LaTeX document, write
%%   \input{<filename>.pdf_tex}
%%  instead of
%%   \includegraphics{<filename>.pdf}
%% To scale the image, write
%%   \def\svgwidth{<desired width>}
%%   \input{<filename>.pdf_tex}
%%  instead of
%%   \includegraphics[width=<desired width>]{<filename>.pdf}
%%
%% Images with a different path to the parent latex file can
%% be accessed with the `import' package (which may need to be
%% installed) using
%%   \usepackage{import}
%% in the preamble, and then including the image with
%%   \import{<path to file>}{<filename>.pdf_tex}
%% Alternatively, one can specify
%%   \graphicspath{{<path to file>/}}
%% 
%% For more information, please see info/svg-inkscape on CTAN:
%%   http://tug.ctan.org/tex-archive/info/svg-inkscape
%%
\begingroup%
  \makeatletter%
  \providecommand\color[2][]{%
    \errmessage{(Inkscape) Color is used for the text in Inkscape, but the package 'color.sty' is not loaded}%
    \renewcommand\color[2][]{}%
  }%
  \providecommand\transparent[1]{%
    \errmessage{(Inkscape) Transparency is used (non-zero) for the text in Inkscape, but the package 'transparent.sty' is not loaded}%
    \renewcommand\transparent[1]{}%
  }%
  \providecommand\rotatebox[2]{#2}%
  \newcommand*\fsize{\dimexpr\f@size pt\relax}%
  \newcommand*\lineheight[1]{\fontsize{\fsize}{#1\fsize}\selectfont}%
  \ifx\svgwidth\undefined%
    \setlength{\unitlength}{121.41733004bp}%
    \ifx\svgscale\undefined%
      \relax%
    \else%
      \setlength{\unitlength}{\unitlength * \real{\svgscale}}%
    \fi%
  \else%
    \setlength{\unitlength}{\svgwidth}%
  \fi%
  \global\let\svgwidth\undefined%
  \global\let\svgscale\undefined%
  \makeatother%
  \begin{picture}(1,0.8411618)%
    \lineheight{1}%
    \setlength\tabcolsep{0pt}%
    \put(0,0){\includegraphics[width=\unitlength,page=1]{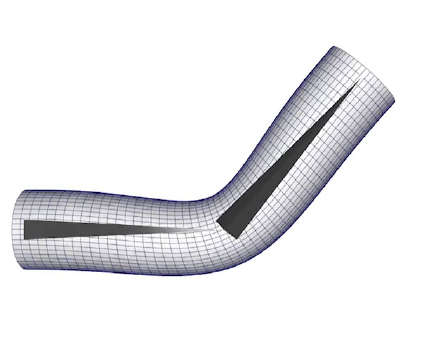}}%
    \put(0.5005446,0.30531454){\color[rgb]{0,0,0}\makebox(0,0)[rt]{\lineheight{1.25}\smash{\begin{tabular}[t]{r}$\PosBi$\end{tabular}}}}%
    \put(0.70321883,0.6870798){\makebox(0,0)[t]{\lineheight{1.25}\smash{\begin{tabular}[t]{c}$\PosVi$\end{tabular}}}}%
    \put(0,0){\includegraphics[width=\unitlength,page=2]{schema.pdf}}%
    \put(0.52475549,0.08176664){\makebox(0,0)[lt]{\lineheight{1.25}\smash{\begin{tabular}[t]{l}$\AngVelBi$\end{tabular}}}}%
    \put(0.48978341,0.61749741){\makebox(0,0)[t]{\lineheight{1.25}\smash{\begin{tabular}[t]{c}$\RotVelViBi$\end{tabular}}}}%
  \end{picture}%
\endgroup%

%\resizebox{0.25\linewidth}{!}{\input{figures/schema.pdf_tex}}
\end{minipage}

\noindent Vector $\TraVelViBi$ 
%(that is, 
(the linear velocity of vertex $\Vi$ induced by the \emph{translation} of bone $\Bi$) is  %just defined as 
equal to the bone's translation $\VelBi$, for all vertices.
While in most character animations, translation is only allowed for the root,  we do not make such assumption in our method, allowing 
translation, eg. when limbs lengths are animated.

\subsection{Procedural velocity-driven deformations} 
\label{sec:deform}

%In the proposed approach, 
In our method, we deform the mesh by 
computing a displacement for each vertex 
expressed as the weighed sum of displacements $\Fun_{\Bi}$ (Equation~(\ref{eq:displacements})), that are function of the vertex velocity component associated with bone $\Bi$. Now that we have selected our weights (Section~\ref{sec:decompose}) and isolated the contribution of each rotational and transitional bone animation (Section~\ref{sec:vel}), let us detail the way we compute $\Fun_\Bi$. 
While the latter takes into account the geometry of bone $\Bi$ (e.g., the location of its origin $\PosBi$),
we will omit the index $\Bi$ in $\Fun$ below,
%in the remainder of the paper, 
for clarity in the exposition.
 \\
To increase the expressiveness of the method
%of individual deformers, 
% MP: "deformers" were not defined yet!
we opt for a version of $\Fun$ that affects the components of Equation~\eqref{eq:velDecompose} differently (rather than being applied to their sum), yielding
\begin{equation}
\label{eq:displacementsBis}
\DisVi = \sum_\Bi{ \AncWeiViBi \, \Fun (\TraVelViBi , \RotVelViBi )  } .
\end{equation}

\noindent Finally, as we want function $\Fun$ to combine 
%an array 
a variety of effects, we set
\begin{equation}
 \label{eq:deformers3}
 \Fun(\TraVelViBi,\RotVelViBi )  = \sum_\text{deform} \Fun_{\text{deform}}(\TraVelViBi,\RotVelViBi ) , 
\end{equation}

\noindent for a number of deformer functions $\Fun_{\text{deform}}$. Thus, to compute $\DisVi$,  all that is left is to define 
%the custom influence of 
the action of each deformer from their constituent velocity terms.  Our squashy and floppy deformers showcase this process in the next section.

\section{Squashy and Floppy Effects} 
\label{sec:squashyFloppy}

While the velocity skinning framework is general and 
%can be applied to any deformation 
can be used for any deformation triggered by the motion of an articulated character, we illustrate it
%apply it 
in this paper through its application to 
%that it can be used to mimic the 
\emph{squashy} and \emph{floppy} expressive effects,
motivated by the classic 
principles of animation~\cite{illusionOfLife}.

\begin{figure}
  \includegraphics[width=\linewidth]{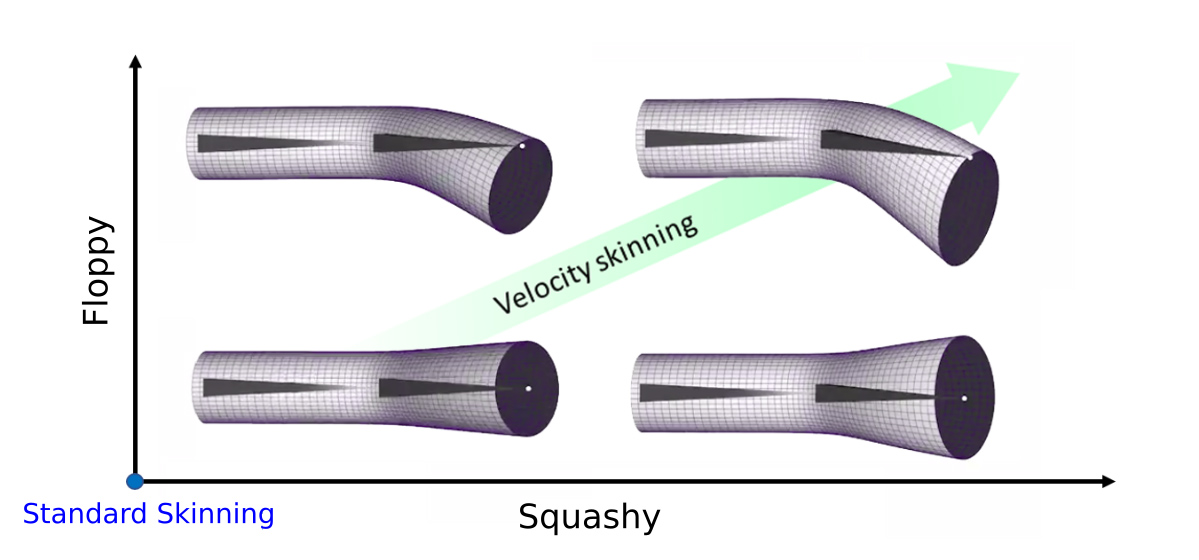}
  \caption{The deformers of squashy and floppy velocity skinning
  reveal information in the still about the speed and direction
  of the motion that is not visible in the standard skinning
  surface. 
  }
  \label{fig:squashflop}
\end{figure}

Both effects can be defined in terms of their translational and rotational input, $\TraFun$ and $\RotFun$, respectively, 
through simple sums of the corresponding terms:
%from the sums of the two  influences:
\begin{equation}
\begin{split}
    \Fun_{\text{squash}} (\TraVelViBi,\RotVelViBi )  &= \TraFunSquash(\TraVelViBi)   + \RotFunSquash (\RotVelViBi ) \\
    \Fun_{\text{floppy}} (\TraVelViBi,\RotVelViBi ) &= \TraFunFloppy (\TraVelViBi) + \RotFunFloppy(\RotVelViBi )
\end{split}
\end{equation}
\noindent with these added to the set of the deformer functions of Equation~(\ref{eq:deformers3}).
The respective scales of these effects can be tuned down (or nullified) using ``floppiness'' and ``squashiness'' gain values, $\PaintSquash$ and $\PaintFloppy$ respectively (see Figure~\ref{fig:squashflop}). These values can be defined per-vertex and ``painted'' on the surface (see Section~\ref{sec:painting}).
\subsection{Squashy deformations} \label{seq:squashy}
Inspired by one of the most well known of animation principles \emph{squash \& stretch}, our squash effect aims to deform an object to produce a local elongation in the direction of motion \cite{illusionOfLife}. At the same time, it is important that this deformation approximately preserves volume.

We define the squash effect through controlled scalings. 
Let the \emph{centroid}  $\CentroidBi$ of bone $\Bi$ (the purple dot in Figure~\ref{fig:SquashyRot}) be the center for the portion of the mesh that is affected by bone~$\Bi$.
Because our system is not physically based, this is not a strict definition, and $\CentroidBi$ can be freely customized by the user (see Section~\ref{sec:manualBoneCenter}). By default, we position $\CentroidBi$ at the barycenter of the vertices in the %affected
bone's region of influence (computed as described in Appendix~\ref{sec:baricente}). Unless stated otherwise, we use this setting in all our examples.
%we adopted this value in all our examples.

\begin{figure}
  \centering
  \includegraphics[width=0.9\linewidth]{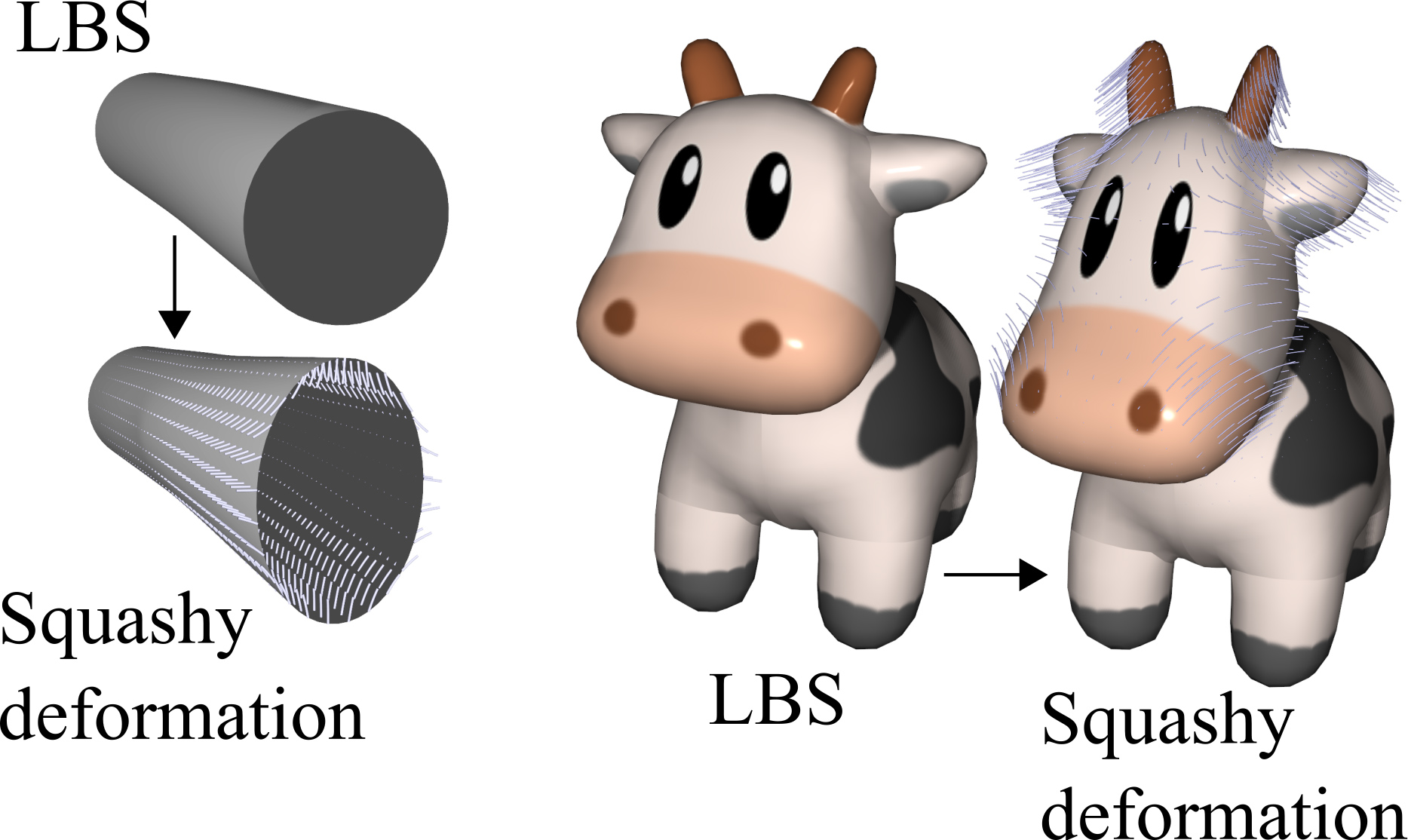}
  \caption{The "squashy" deformer applies an anisotropic scaling to the moving parts of the shape compared to the shape obtained by LBS. Left: Result when the front-part of the cylinder rotates downward; Right: Result when the cow's head rotates downward.}
  \label{fig:squashy_deformation}
\end{figure}

\begin{figure}
  \centering
  \includegraphics[width=1.0\linewidth]{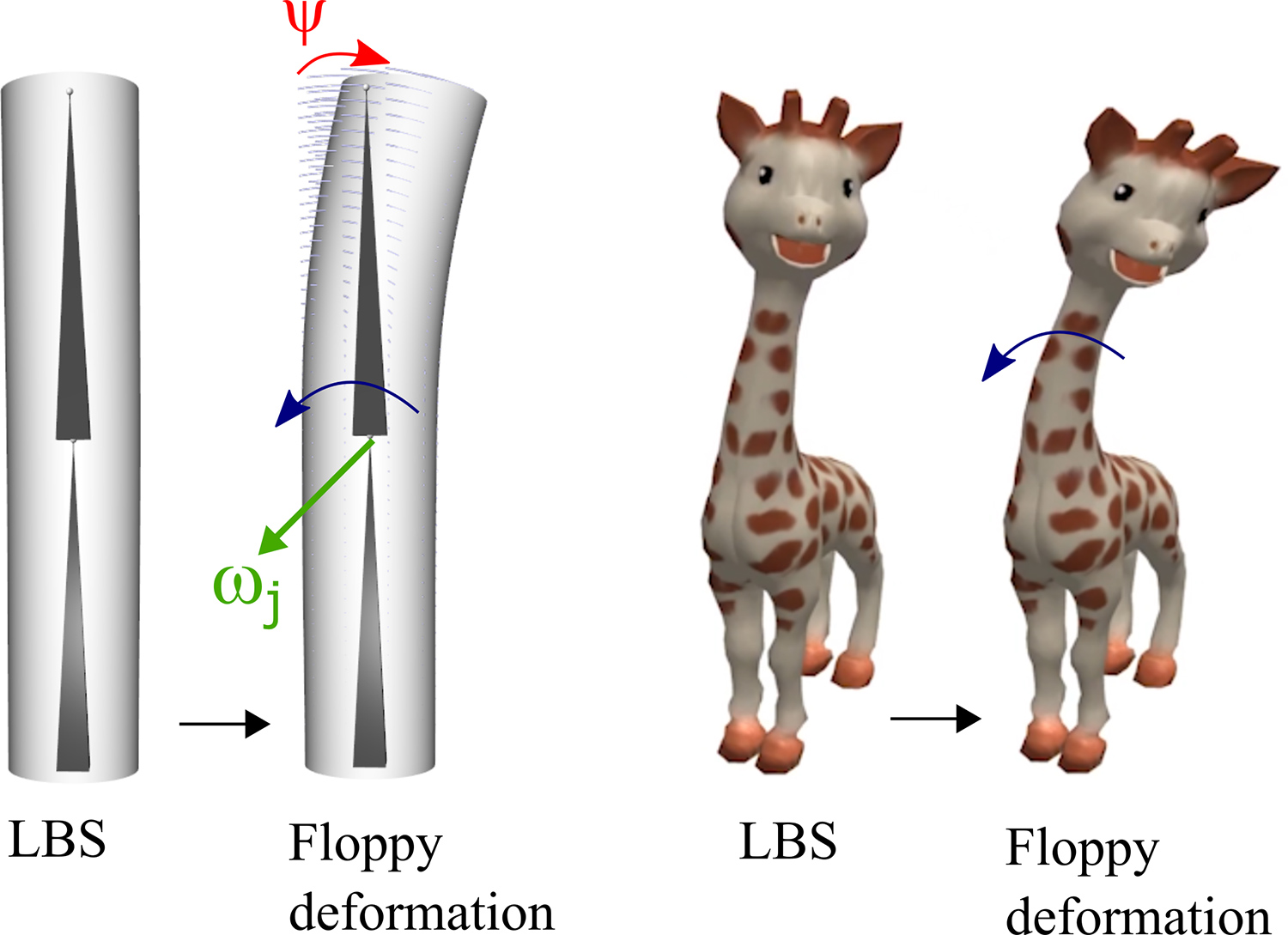}
  \caption{Left: "Floppy" deformation induced by the counter-clockwise rotation of the top bone. In rest pose, the shape is a straight cylinder, while the deformer bends the cylinder in the opposite direction. Right: Same deformer 
 % applied to a 
  when a neck bone of the giraffe rotates.}
  %The ``floppy'' deformation induced by the counter-clockwise rotation of the right-side bone. In rest pose, the shape is a straight cylinder.
  \label{fig:bendyarm}
\end{figure}

\emph{For linear bone motions,} the scaling is simply centered in  $\CentroidBi$. The formula for the displacement is given by

\begin{equation}
\label{eq15}
\TraFunSquash (\TraVelViBi ) = (\matrix{R} \, \matrix{S} \, \matrix{R}' \, - \, \matrix{Id} ) (\PosVi - \CentroidBi ) \,,
\end{equation}
where \(\matrix{Id}\) is the identity matrix, $\matrix{R}$ is any rotation matrix which maps the $x-$axis to the direction of the bone velocity $\VelBi$ (and $\matrix{R}'$ denotes its transpose); $\matrix{S}$ is the following anisotropic, volume-preserving scaling matrix:
$$
\matrix{S}=
\begin{bmatrix}
1+s&0&0\\ 0&1/\sqrt{1+s}&0\\ 0&0&1/\sqrt{1+s}
\end{bmatrix}
$$
\noindent where 
the scaling value $s$ is proportional to the speed of the velocity component:
$$s = \PaintSquash \,\| \TraVelViBi \| $$

\emph{For rotating bone motions,} the scaling resembles the effect of a centrifugal force generated by a spin. In this case, for each bone $\Bi$, we identify a \emph{medial axis}, as, intuitively, an axis expected to run approximately through the interior of the shape; as a default, 
we define it as the axis connecting $\CentroidBi$ with $\PosBi$ (magenta line in Figure~\ref{fig:SquashyRot}), but again this choice can be tuned.
Then, we want the squash effect not to produce any elongation or shrinking along this axis. 
Let $Pr$ be the operator projecting a point into this medial axis. 
The formula for the displacement is then given by
\begin{equation}
\label{eq:RotFunSquash}
\RotFunSquash (\RotVelViBi ) = (\matrix{R} \, \matrix{S} \, \matrix{R}' \, - \, \matrix{Id} ) (\PosVi - Pr(\PosVi)  ) \,,
\end{equation}
\noindent where $\matrix{R}$ is the rotation that maps the $y$-axis parallel to medial axis, and maps the $z$-axis as close as possible to $\AngVelBi$ %(the construction of this matrix is in Appendix~\ref{sec:deriveR}). 
When $\AngVelBi$ is parallel to the medial axis, then $\matrix{R}$ is undefined, and  $\RotFunSquash$ is simply zeroed.  $\matrix{S}$ is the following anisotropic, volume-preserving scaling matrix:
$$
\matrix{S}=
\begin{bmatrix}
1+s&0&0\\ 0&1&0\\ 0&0&1/(1+s)
\end{bmatrix}
$$
\noindent where the scaling value $s$ is proportional to the magnitude of of the relevant velocity component:
$$s = \PaintSquash \,\| \RotVelViBi \| $$

\subsection{Floppy deformations} \label{sec:floppy}

The floppy deformer is inspired by an exaggeration effect described by classic animators~\cite{illusionOfLife} as ``\textit{The loose flesh [...] will move at a slower speed than the skeletal parts. This trailing behind in an action is sometimes called `drag', and it gives a looseness and a solidity to the figure that is vital to the feeling of life.}'' which is taken from a passage about their rule for Follow Through and Overlapping Action.
We adopt the term ``floppy'' to describe this effect. Intuitively, this effect stems from the skin vertices appearing to move with a delay (or \emph{drag}) with respects to the initiating bone movement; we obtain this by displacing vertices opposite to the velocity induced by each  individual bone.

\emph{For linear bone motions,}
this is achieved by simply displacing the vertex in the opposite direction of the linear velocity induced by bone $\Bi$:
\begin{equation}
    \TraFunFloppy( \TraVelViBi ) = -\PaintFloppy \, \TraVelViBi
\end{equation}
with the ``floppiness'' value, $\PaintFloppy$, reflecting how pronounced the effect must be (this value can be defined per vertex, see Section~\ref{sec:painting}).

\begin{figure}
  \centering
  \includegraphics[width=1.0\linewidth]{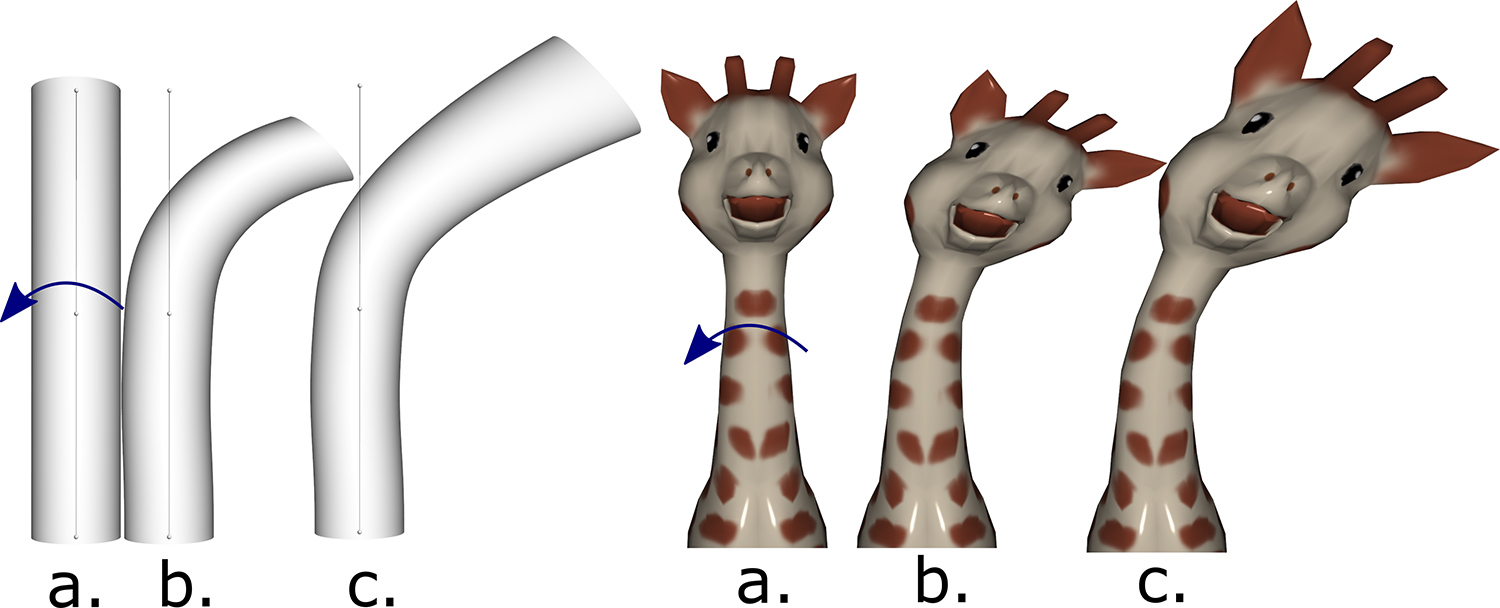}
  \caption{\final{
  For a tube (left) and the giraffe model (right), (a.) is the initial shape with an arrow indicating motion. The drag deformation using velocity-skinning (b.) adequately models bending, while the use of per-vertex velocities produces a stretching effect (c.).
  %a. Initial shape and associated rotation motion; b. Drag deformation using our velocity-skinning formulation models bending; c. Drag generated from the per-vertex velocity vector is associated to stretching effect.
  }}
  \label{fig:vertex_velocity}
\end{figure}

\emph{For rotating bone motions,}
 the delay produces a bending (rotation) around the current axis rotation of bone $\Bi$ (that is, the axis passing through $p_i$ and aligned with $\omega_i$), in the opposite direction:
\begin{equation}
    \RotFunFloppy( \RotVelViBi ) = (\matrix{R}-\matrix{Id} ) (\PosVi-Pr(\PosVi) )
    \label{eq:RotFunFloppy}
\end{equation}
where $Pr$ is the operator projecting $\PosVi$ on the rotation axis, and $\matrix{R}$ is the rotation matrix around said axis with angle
$$
\label{eq:angle}
\theta = - \PaintFloppy \, ||\RotVelViBi|| \,.
$$
Crucially, as $||\RotVelViBi||$, and thus $\theta$, increase linearly with the distance to the bone (as per Equation~\eqref{eq:joint}), the displacements induced by Equation~\eqref{eq:RotFunFloppy} is not linear, resulting in naturally looking deformations that curve the profiles (as exemplified in Figure~\ref{fig:bendyarm}).
\final{In addition, the advantage of defining our deformation from the angular velocity of the joint, instead of using per-vertex velocities, is the ability to bend the shape, even by large angles, without introducing unwanted stretch (see  Fig.~\ref{fig:vertex_velocity}}).
%We further note that the advantage defining a deformation based the angular velocity component of the joint instead of the per-vertex velocity is the ability to bend the shape, even for large angle without introducing unwanted stretch as illustrated in Fig.~\ref{fig:vertex_velocity}}.

\section{Deformer Controls}\label{sec:interface}

In this section we introduce interactive mesh deformation tools which allow simple customization of parameters 
with simultaneous visualization of their effect on the character.

\begin{figure}
  \includegraphics[scale=0.13]{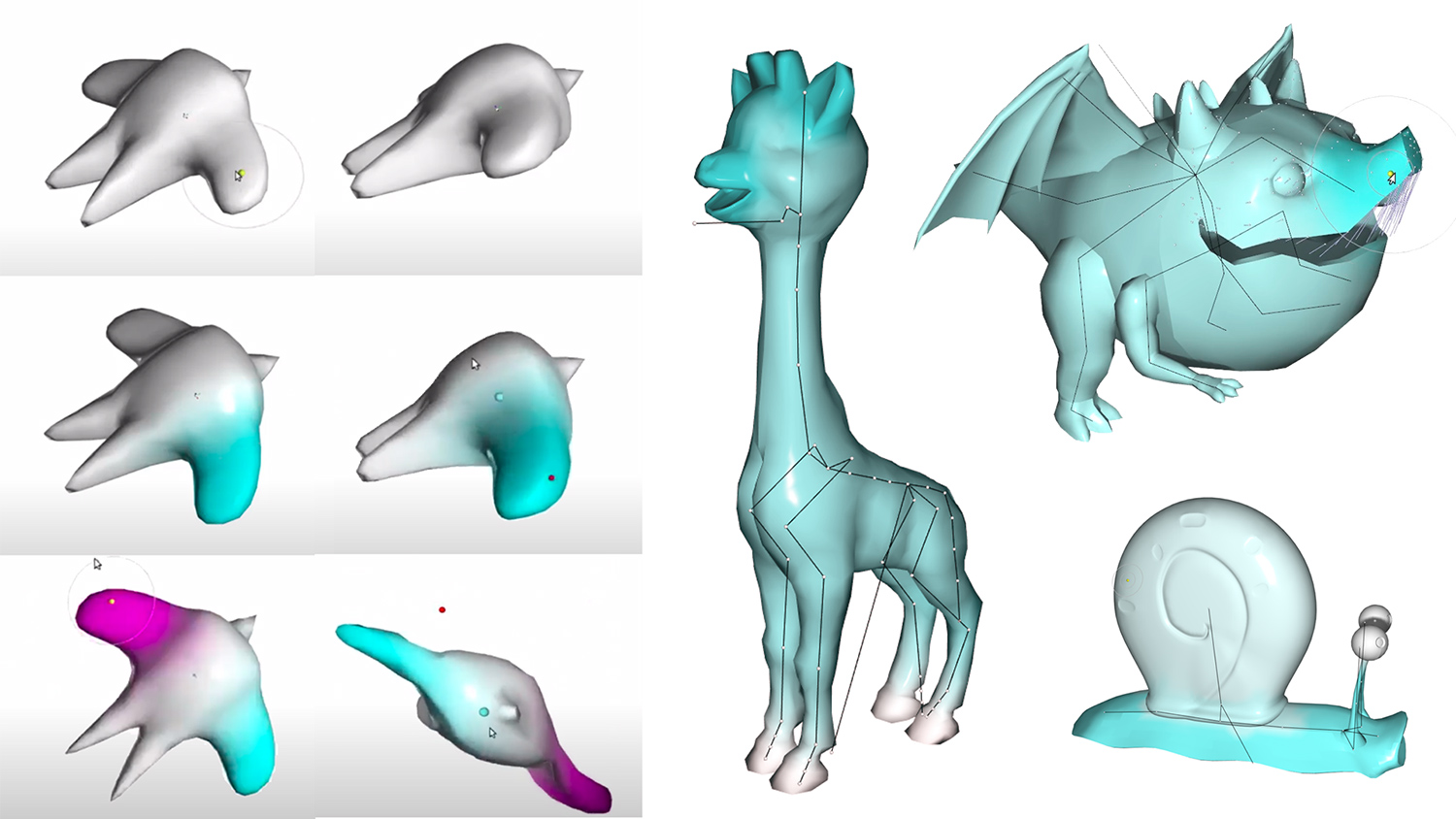}
  \caption{Weight painting: left column, painting interface allows positive (cyan) and negative (magenta) weights to dictate the influence of the skinning effects. Left center column, representative image after weights are applied.  Note top row equivalent to standard skinning with no effects on the outcome.  Right, example paintings
  that show weights in practice on a three different models.}
  \label{fig:Weight}
\end{figure}

\emph{Per-vertex painting.}\label{sec:painting}
$\PaintFloppy$ and $\PaintSquash$ weights (see Section~\ref{sec:squashyFloppy}) can be set non-uniformly over the surface in defining them as a per-vertex scalar value. These weights can be interactively painted over the surface using a readily available brush-like tool in standard modeling software.
Non-uniform weights allow the user to express local effects that could not necessarily be captured by the animation skeleton.
Auxiliary "phantom" bones (a.k.a. helper bones) are not required, we just need to set high weights values to make parts of the mesh look softer, or zero weight values to make them look rigid. Non-uniform weights can bring stylization effects even in the extreme case of a mesh associated with a single joint. 
%\MP{
%Such example is provided 
This is illustrated by the bird in Figure~\ref{fig:Weight}-left, experiencing a rigid translation of its single joint, positioned at its center. The non-uniform floppy-deformation weights used for the wings (blue color) locally enhance motion.
%related to the bone's translation.
%}
An extra degree of freedom can be used in associating negative values to weights -- thus making deformation occur in \emph{anticipation} of bone motion, instead of having skin parts \emph{drag} behind. Lastly, the combination of positive and negative weights enables either synchronized motion or may provide a sense of action/reaction such as a birds flapping its two wings in one direction and in the opposite one (see bird wings in the accompanying video and in the interactive online demonstration). 

\begin{figure}
  \includegraphics[width=0.9\linewidth]{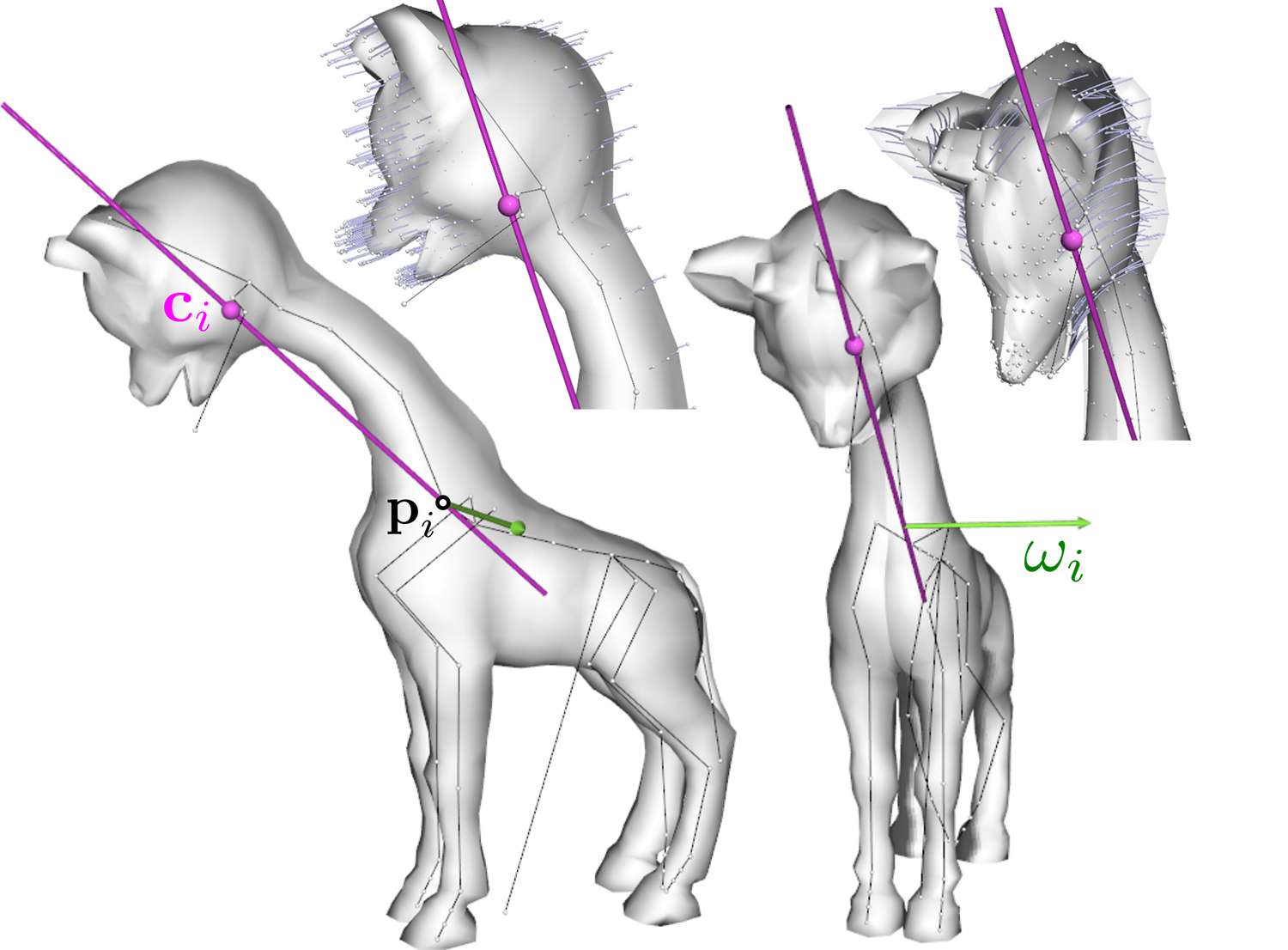}
  \caption{Squashy deformer applied to the neck of a giraffe model. The magenta dot is the \emph{centroid} $\CentroidBi$ for the lowest neck bone, and the magenta line, its \emph{medial axis}. These artist-tunable parameters are the centers for translation and rotation, for the squash effect.} % I (Faezeh) removed the word "respectively" after rotation.
  %\vspace{-1em}
  \label{fig:SquashyRot}
\end{figure}

\emph{Per-bone control.}\label{sec:bone-control}
A coarse-level controller is added to manage which bones are affected by each effect.  In
this manner, entire branches can be removed (e.g. the legs in a walking motion) to assure
the intent of the keyframe animation is not impacted negatively.  Further, the bones that
receive the effect can be controlled with finer granularity, by setting the weights
of all bones, separately.  This includes a split of the rotational and translational 
components of each.  The artist supplemental video highlights more of the distinctions 
that can be made through this level of control.

\emph{Squashy-specific controls.} \label{sec:manualBoneCenter} 
We provide interactive control for the squash effect, in allowing the user to interactively move the centroid \(\CentroidBi\) associated with a bone (see Section~\ref{seq:squashy} and Figure~\ref{fig:SquashyRot}). This enables a fine tuning of the center or the axis around which the elongation and compression are applied for each bone, similar to the scale-pivot interface of standard 3D packages. The user-displacement applied to \(\CentroidBi\) is stored as an offset expressed in the local coordinate frame of the bone, and is therefore automatically adjusted during the animation. In addition, the user can force the squashy effect to act with respect to a single position instead of an axis. While the squash effect around an axis is well adapted for rotating limbs, the user may rather apply the effect of end-effector bones with respect to the local centroid
in specific cases.

\begin{figure}[h!]
\includegraphics[width=\linewidth]{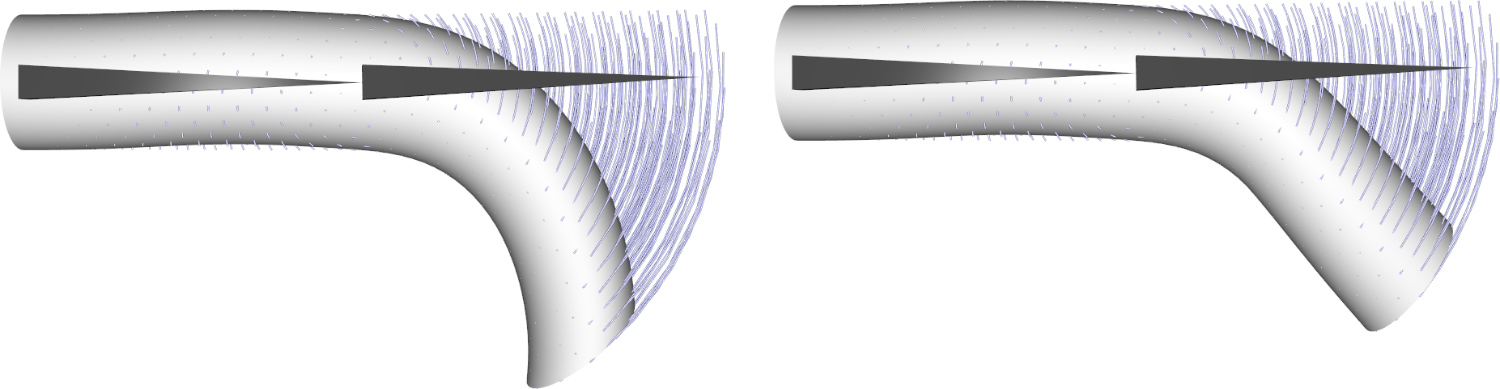}
\caption{
Deformation of a straight cylinder, when the right-bone rotates counter-clockwise. Left: High speed or large floppy weights may lead to exaggerated bending deformation. Right: A maximal bending angle (in this case $45^{\circ}$) is set to bound the deformation. %Note that the LBS shape is a straight cylinder, and that the right-bone rotates counter-clockwise.
% I (Faezeh) removed the word "overly" before exaggerated
}
\vspace{-1em}
\label{fig:max-angle}
\end{figure}

\emph{Floppy-specific controls.}
The floppy effect is similar to the surface-bending effect in some 3D modeling packages.
However, ours is automatically scaled with the speed (to add the appearance of dynamics effects) and the floppy weight values (to give higher controllability). When high speed animation is used, this bending deformation may get exaggerated above plausible (desired) range. We integrate a simple interactive limiter defined as a maximal allowed bending angle for \(\theta\) from Section~\ref{sec:floppy}, illustrated in Figure~\ref{fig:max-angle}. This allows the user to model salient floppy effects for a character exhibiting both slow and high speed motions, while ensuring that high speed related deformations remains in the admissible range.

\begin{figure}[h!]
\includegraphics[width=0.9\linewidth]{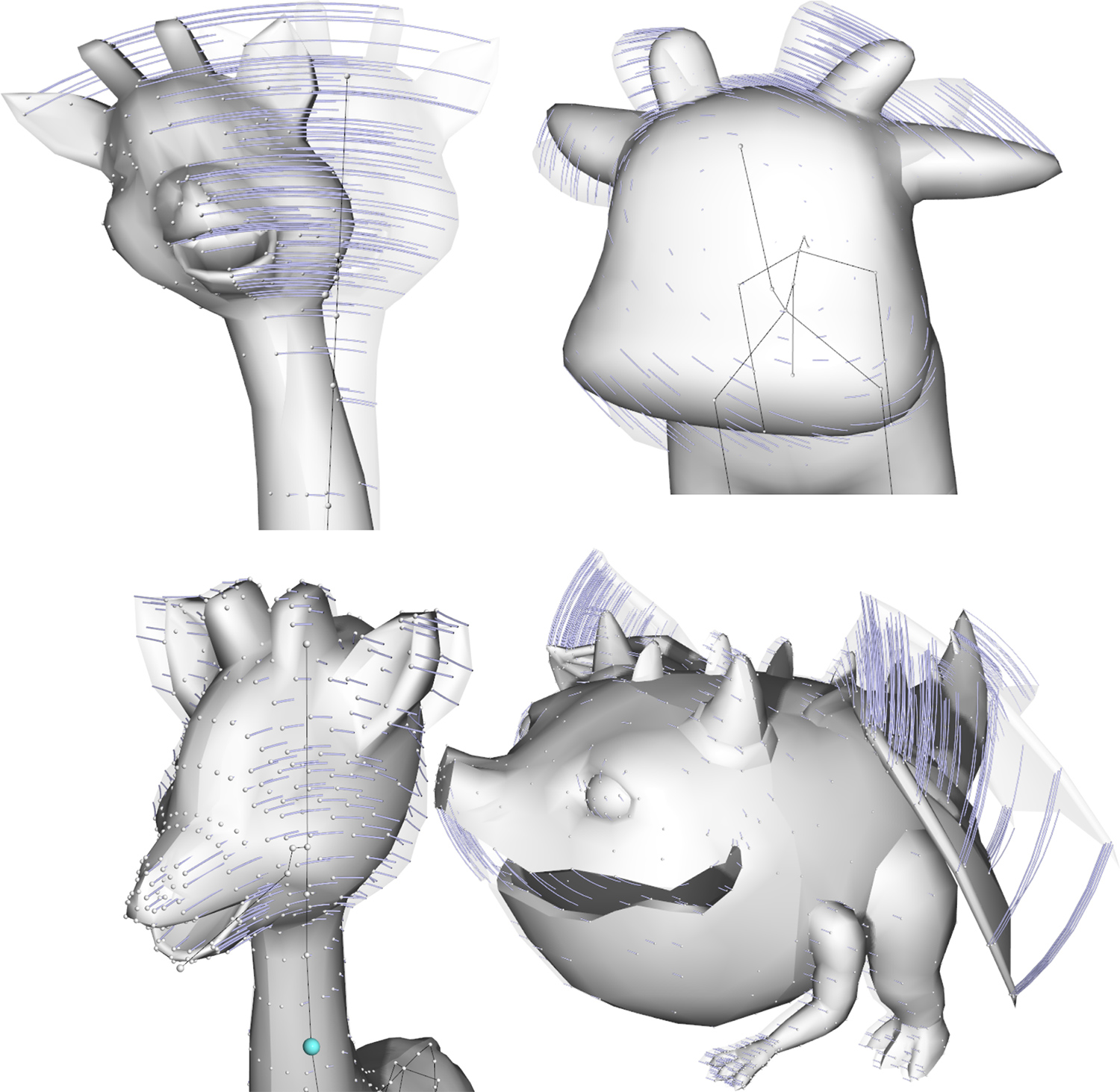}
\caption{
Trajectory followed by skin vertices moving from standard LBS (depicted as a transparent surface), to 
%ward the shape obtained after 
\textit{velocity skinning} deformation (opaque surface). 
Top: Floppy effect after bending the giraffe neck (left), and twisting the cow head (right). Bottom: Squashy effect after twisting the giraffe neck (left), and multiple floppy deformations applied on different parts of the dragon (right). }
\label{fig:trajectory}
\end{figure}

\emph{Deformation Visualizer.}
%\emph{Freeze-Velocity Deformation Visualization.}
Interactive tuning of time-dependant deformation can be a complex task, while fine tuning is easier in static poses. The deformations we generate, however,  fundamentally rely on velocity, although not directly on a time variable. Therefore, constant linear and angular velocity values can be artificially attached to bones, allowing a visualization of the induced deformation in a static pose/velocity space. The velocity (as well as the different parameters described previously) can then be interactively modified, while the deformation is updated accordingly. To ease the visualization of the deformation with respect to the input skin mesh, as well providing insight about the underlying motion and speed, our visualizer traces the trajectory that vertices would follow when the velocity magnitude of bones increase from 0 to the set values (see Figure~\ref{fig:trajectory}). This tool offers an enhanced, visual understanding of the modeled deformation, independent from the final skeletal animation, and provides visual cues enabling tuning of velocity skinning effects from a static pose.

%\section{Implementation} \label{sec:implementation}
\section{Data flow of Velocity Skinning} \label{sec:implementation}  
\label{sec:algorithm}

Velocity skinning is well suited to standard real-time animation pipelines, since it follows the same setup both at pre-processing and at run-time stages.
In particular, the independence between animation steps is well preserved, as all the data needed at run-time (in addition to skinning) can be stored separately for each skinned model (independently from the animations), and for each skeletal animation (independently from the models).

\emph{Pre-processing a skinned model.}
At this stage and for each model, we propagate original bone weights along the skeleton structure to compute velocity weights using Equation~(\ref{eq:AncWei}), and store them per vertex. User-defined \emph{drag} bounds and weightings (see Section~\ref{sec:painting}) are also stored as vertex attributes of the skinned model. We also compute the centroid of each bone (Equation~(\ref{eqBarycenters})), and store the manually tuned offset the user may add (see Section~\ref{sec:manualBoneCenter}).

\emph{Pre-processing a skeletal animation.}
Given an animation (defined as a skeleton pose per keyframe) per-bone angular and linear velocities are extracted using derivatives of the joint trajectory curves as explained in Section~\ref{sec:vel}.
Our approach can accommodate arbitrary animation inputs such as parametric key-framed animations, procedural motions, as well as interactive run-time manipulations. To avoid discontinuity artifacts coming from sudden changes of position, these velocities can be smoothed-out by time averaging.
%This applies to animations computed procedurally, such as, for example, rag-dolling techniques. If necessary, the velocity can be smoothed-out by time averaging, to avoid discontinuity artifacts. To accommodate run-time user manipulation, these velocity pre-computations can be replaced by finite differences \Damien{computed on-the-fly on joint position and orientation} between the previous and current frames.

\emph{At run-time.} After per-animation and per-model data are prepared or generated, final deformations are computed on-the-fly thanks to their closed-form formulation.
In contrast to a physically-based approach, this fully kinematic procedural approach requires no book-keeping of previous states, resulting in a single-pass method, fully applied at run-time. In
our experimentation, we applied velocity skinning on top of two standard skinning approaches, LBS and DQS, and it has a comparable memory footprint and workload to non-velocity skinning implementations.

%\begin{table}
%\begin{tabular}{l|p{1.5cm}|l|r}
%    Model & Velocity Skinning & LBS & \#vertices \\
%    \hline
%     Bird & 0.6\,ms & 0.3\,ms & 604 \\  
%     Giraffe & 1.1\,ms & 0.4\,ms & 1832 \\  
%     Cylinder & 4.0\,ms & 1.2\,ms & 2500\\
%     Cow & 1.6\,ms & 0.9\,ms & 3225 \\  
%     Dragon~1 & 5.5\,ms & 2.4\,ms & 3776 \\  
%     Dragon~2 & 6.5\,ms & 2.4\,ms & 3776  \\ 
%     Snail & 10\,ms & 4\,ms & 8223 
%\end{tabular}
%\label{timing}
%\caption{Timing (in ms) of velocity skinning compared to LBS for different models. All models (including Dragon~1) have only one bone in motion (including the root impacting the entire shape) while Dragon~2 has 7 bones in motion. \Damien{Meaningless since GPU version}}
%\end{table}

\section{Implementations and Results}

\begin{figure*}[th]
  \includegraphics[width=\linewidth]{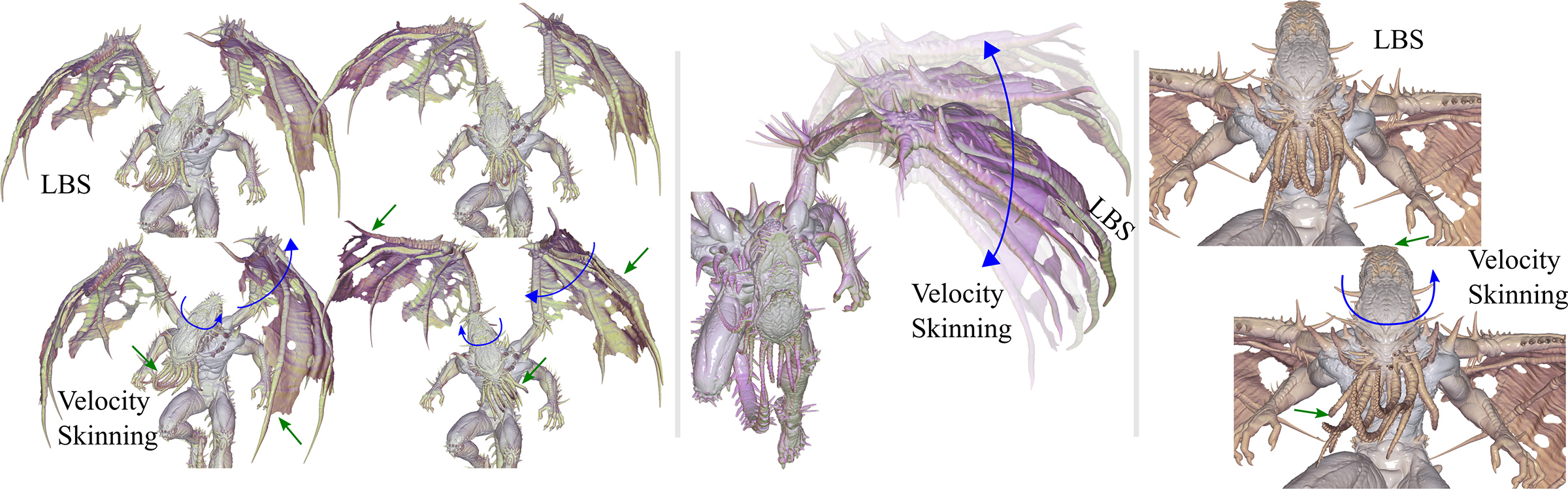}
  \caption{
  Velocity skinning applied on a detailed mesh, 
  thanks to our GPU implementation. The blue arrows indicates the main motions, while the green arrows highlight some noticeable effects of our deformers. Left: Comparison of LBS (top) and velocity skinning (bottom) on two extreme poses when constant floppy weights are used. Velocity skinning automatically enhances the motion in bending the wings and the tentacles based on their respective bones motion. Middle: Shape variations of the wing applied by the floppy deformer, the "neutral"-LBS pose being the middle one. Right: Computing automatic per-vertex weights based on local shape diameter allows to apply more deformation on mesh details, thus mimicking a lower stiffness on small protruding elements.}
  \label{fig:cthulu}
\end{figure*}

\begin{table*}[ht]
    \centering
    \begin{tabular}{r||c|r|r|c||r|r||r|r}
         \multicolumn{1}{c||}{\multirow{4}{*}{Scene}} &
         \multicolumn{4}{c||}{\multirow{2}{*}{Complexity (number of)}} & 
         \multicolumn{2}{c||}{GPU} & 
         \multicolumn{2}{c}{Computing time} \\ 
         &
         \multicolumn{4}{c||}{} &
         \multicolumn{2}{c||}{memory} & 
         \multicolumn{2}{c}{(per frame)} \\ \cline{2-9}
         &
         \multicolumn{1}{c|}{\multirow{2}{*}{Models}} &
         \multicolumn{3}{c||}{per model} &
         \multicolumn{1}{c|}{\multirow{2}{*}{LBS}} &
         \multicolumn{1}{c||}{\multirow{2}{*}{VS}} &
         \multicolumn{1}{c|}{\multirow{2}{*}{LBS}} &
         \multicolumn{1}{c}{\multirow{2}{*}{VS}}  \\ 
         &
         &
         \multicolumn{1}{c}{vert} &
         \multicolumn{1}{c}{tri} &
         \multicolumn{1}{c||}{bones} & 
         &
         &
         &
          \\ \hline \hline 
        %
        %        \emph{\# models} & \emph{\# vert} & \emph{\# tri} & \emph{LBS only} & \emph{VS-full} & \emph{LBS only} & \emph{VS-1-bone} & \emph{VS-full} \\ \hline 
        \multirow{6}{*}{Flying Cthulhu} &
         \multirow{2}{*}{1} & \multirow{2}{*}{150k} & \multirow{2}{*}{300k} & 1 & 
         \multirow{2}{*}{13 MB} & \multirow{2}{*}{21 MB} & \multirow{2}{*}{3.9 ms} & 8.0 ms  \\ \cline{5-5}\cline{9-9}
        & & & & 12 & 
          &  &  & 8.7 ms  \\ \cline{2-9}
         &
         \multirow{2}{*}{1} & \multirow{2}{*}{250k} & \multirow{2}{*}{500k} & 1 & 
         \multirow{2}{*}{22 MB} & \multirow{2}{*}{35 MB} & \multirow{2}{*}{6.6 ms} & 13.9 ms  \\ \cline{5-5}\cline{9-9}
         & & & & 12 & 
          &  & & 14.4 ms  \\ \cline{2-9}
         &
         \multirow{2}{*}{1} & \multirow{2}{*}{750k} & \multirow{2}{*}{1500k} & 1 & 
         \multirow{2}{*}{64 MB} & \multirow{2}{*}{106 MB} & \multirow{2}{*}{18.7 ms} & 34.8 ms  \\ \cline{5-5}\cline{9-9}
         & & & & 12 & 
          &  &  & 36.7 ms  \\ \hline
        
         \multirow{2}{*}{Cow's meadow}&
         \multirow{2}{*}{4000}&
         \multirow{2}{*}{3.2k}&
         \multirow{2}{*}{5.8k}&
         \multirow{2}{*}{7}&
         \multirow{2}{*}{1.7 MB}&
         \multirow{2}{*}{2.4 MB}&
         \multirow{2}{*}{17.2 ms}&
         \multirow{2}{*}{39.0 ms}  \\ 
         & & & & &  &  &  &   \\ \hline
 
        %150k & 300k & 11 MB & 20 MB & 3.9 ms & 5.7 ms & 6.7 ms & \\ \hline
        % 250k & 500k & 19 MB & 33 MB & 6.6 ms & 10.0 ms & 11.4 ms & \\ \hline
        % 750k & 1500k & 56 MB & 104 MB & 18.7 ms & 27.2 ms & 30.9 ms & \\
    \end{tabular}
    \caption{Timings and memory usage measured on different scenes, animated on GPU with standard Linear Blend Skinning (``LBS'') and  Velocity Skinning (``VS''). GPU memory includes the storage all precomputed parameters. 
    %The timings drawing call of the model for 1-frame. ``LBS only'' corresponds to the implementation where all velocity-skinning related parts are commented out, while ``VS'' indicates the use of velocity skinning deformers. ``VS-1-bone'' corresponds to a measure where only one wing of the model is animated. ``VS-full'' corresponds to the complete animation with 12-animated bones. 
    Average timings are shown, but minimal and maximal stay in a range  of \(\pm 0.5ms\). Bones indicates the actual number of bones that have a non-zero motion - thus generating a Velocity Skinning deformation on the mesh. }
    \label{tab:timings}
\end{table*}

%
%This variety of examples showcase the strengths and features of our approach,

For testing purposes, we implemented Velocity Skinning in three working prototypes, which were used to produce all the rendering in the figures in this paper
%(e.g. \ref{fig:teaser} and \ref{fig:Dragon}) 
and in the accompanying videos. The source code of all three prototypes is provided in the supplemental material.

%\MP{\em MP: The two implementations - CPU and GPU -should be discussed in Section 6, not in Section 7 (or change section titles and double-check references to sections in he Introduction?)} \Damien{I changed the title of "Implementation" to "Data processing" as it mostly discuss how data are prepared and handle, rather than the actual implementation. To be validated by somebody.}
 The first implementation is a stand-alone, CPU-based desktop application, serving as a reference implementation for Velocity Skinning. It displays Velocity-Skinned animations, and doubles as an tool for authoring Velocity-Skinned meshes, by allowing an user to set the controlling parameters for a given model, and see the resulting deformation with with a given animations. 
 
%The source code is %available at 
%provided in the supplemental material.
%\TODO{insert github link}.
% Not yet - still anonymous
 % Damien: I say it in the beginning - all source code is provided
 
%\sout{We performed some stress testing of our system in a limited manner by having a novice and (more) trained animator produce animations, namely a researcher on our team and a student training as animator.
%Each produced one of the two cow animations that bookend the video accompanying this paper.  Both exhibit a modest quality of stylized effects, after very little training.
%While further testing is clearly desirable, these early experiments provide a humble
%first validation of the effectiveness of our method. We discuss more on user testing, along with other findings and limitations next.}
 
\begin{figure*}
  \includegraphics[width=\linewidth]{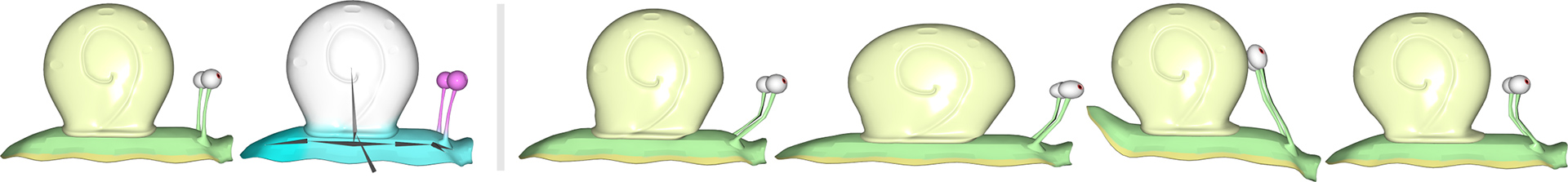}
  \caption{Example of deformations generated using time-varying skeletal velocities from a single skeleton pose. Floppy weights are shown in the second image: zero on the shell; negative weights on the eyes; and positive weights on the rest of the body. The squash effect is applied uniformly on all parts of the shape.
  }
  \label{fig:snail_history}
\end{figure*}

In order to explore the range of effect which can be achieved,
 and as a preliminary test of the authoring interface, 
we solicited the assistance of a set of animators: two novices (first-year digital-art graduate students) and two professionals (with 5-10 years experience
in the commercial sector). 
The animations appear within the primary video along with the representative 
research results, but the process of adding effects is also highlighted in a supplemental video with side-by-side animations that showcase the input and 
output of our system.  This video is narrated by one of the (first-year) animators describing the addition of effects in a ``tutorial'' style.  The findings suggest our tools can add controllable effects to animations from both categories, novice and professional.  Namely,
the novice animations benefited from adding coarse effects (e.g. the whole head
of the cow model) while the more refined professional animations only needed small,
more subtle additions to push the existing animations further.  An example of the
latter would be using weight painting to isolate floppy effects to the ears and tail of the cow which do not include rig-bones in our stylized model (see Figure~\ref{fig:teaser}-left). Another example is illustrated in Figure~\ref{fig:snail_history} by a snail with different weights on its respective eyes, body and shell parts.
%For instance, it allows to swap between two versions of a skinned ``snail character'' models, which differ on the ``squishiness'' of the shell part (see  Figure~\ref{fig:snail_history}).  %
In a post interview, the (professional) animator of the walk-cycle animation stated that it would take him an approximated three hours in Maya, based on his experience, to add the effects we introduced through our tools.  It took a less-skilled animator less than an hour to produce the shown result.
See the video for more details of this hands-on experimentation.

%\TODO{Probably say something about how this experience went. How happy the test artists have been, what did they do.}

The second implementation is an interactive web-based demo that displays animations on a small set of character models, intended to showcase the effects that can be achieved with Velocity Skinned. It enables users to compare Velocity Skinning with standard blend skinning, and to test different pre-made set of parameters, which are customized for
specific characters. The web-demo is based on WebGL, but computes the deformation on the CPU side. It can be accessed at \href{https://velocityskinning.com}{https://velocityskinning.com}.
%\TODO{insert web-demo URL here}

The third implementation is a fully GPU-based desktop application, implementing Velocity Skinning in a vertex shader. It is intended to empirically evaluate the scalability of Velocity Skinning in a video-game or similar context. In this implementation, the different pre-processed parameters, including rig-weights and its propagated version as well as skeleton poses, are sent to the GPU memory as buffer objects, and accessed in the shader. 
%Note that rig weights (\(\WeiViBi\) and \(\AncWeiViBi\)) were not restricted to any maximal number of bone dependency, so the full indexed-connectivity of each vertex was therefore sent on the GPU memory.
%
%
%

We tested this system on different scenarios: scenes composed by a single large model, with 300k, 500k and 1.5 millions of triangles (three resolution levels of a ``Cthulhu'' model~\cite{cthulhu}) illustrated in Figure~\ref{fig:cthulu}, and a scene with 4000 individually animated instances of a low-res "cow" models of 3200 vertices each (see Figure~\ref{fig:teaser}-right). To isolate the impact of the number of bones, we also compared the performances with the large model using only one animated bone. In each case, we compare against the case of Linear Blend Skinning. Table~\ref{tab:timings} reports total GPU memory usage, and render times measured on a consumer laptop in all scenes (GPU: NVIDIA Quadro P3200; CPU: Intel Core i7 2.6GhZ CPU).

This experiment indicates that velocity skinning can easily fit to low budget computation, allowing for real-time rendering even in presence of high-resolution mesh, and complex scenes. 

%This shows that velocity skinning is comparable to basic skinning in terms of performance, while the method does add cost as vertex-to-bone links are less sparse in general, due to propagation to ancestor bones (Equation~\eqref{eq:AncWei}), and because per-bones velocities need to be extracted or baked for an animation. 
%Even in such cases, computational times still fit within requirements of most real-time applications.

%
\begin{figure}
  \includegraphics[width=\linewidth]{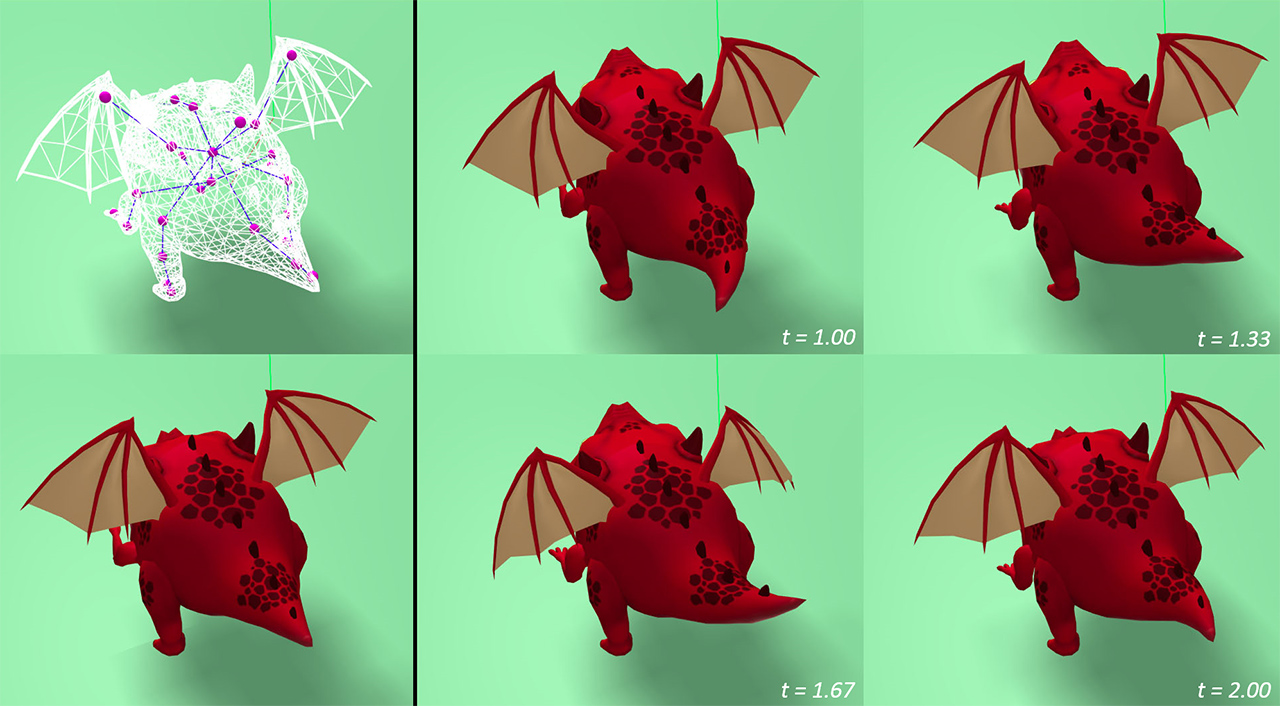}
  \caption{This dragon has a minimal number of bones in its wings and tail, left column shows the rest post and skeleton.  Through velocity skinning, a rich flapping and wagging animation (right, frames) can be added automatically to augment the limited skeleton animation.  See accompanying video for more detail.}
  \vspace{-2em}
  \label{fig:Dragon}
\end{figure}

\section{Discussion and Conclusion} \label{sec:discussion}

We introduced a new technique, velocity skinning, which enriches standard skinning with automatic stylization, without significantly impacting computational times. The main idea is to add a displacement at each vertex, computed from the velocities of the bones with non-zero local LBS weights.
We demonstrated the usefulness of the method with two specialized deformers generating squashy and floppy deformations, designed in accordance with two classic animation principles. 
The resulting technique is finely controllable via a set of easily edited controls, such as ``painting'' the extent of the individual effects on the 3D model.

A key of the simplicity and thus efficiency of velocity skinning is the additive nature of the applied displacement (i.e. all the procedural displacements derived by individual bone motions, across all stylization effects, are simply added up).
This is justified by the central observation that vertex velocities can be decomposed as a weighted sum of skeletal influences. 
However, because the deformation functions $\Fun$ are in general not linear with velocity, this is only an approximation which degrades as deformations grow.
Thus, velocity skinning cannot be expected to always give good results, especially for extreme deformations or special conditions, for example when mixing large deformations induced by velocities with different directions.
In spite of this, our experiments indicate that velocity skinning is able to produce expressive stylized animations under many useful scenarios, such as when the skinned mesh is animated by a rig with only a few joints (see Figure~\ref{fig:Dragon} or the web demonstration).

\begin{figure}[th]
  \includegraphics[width=0.95\linewidth]{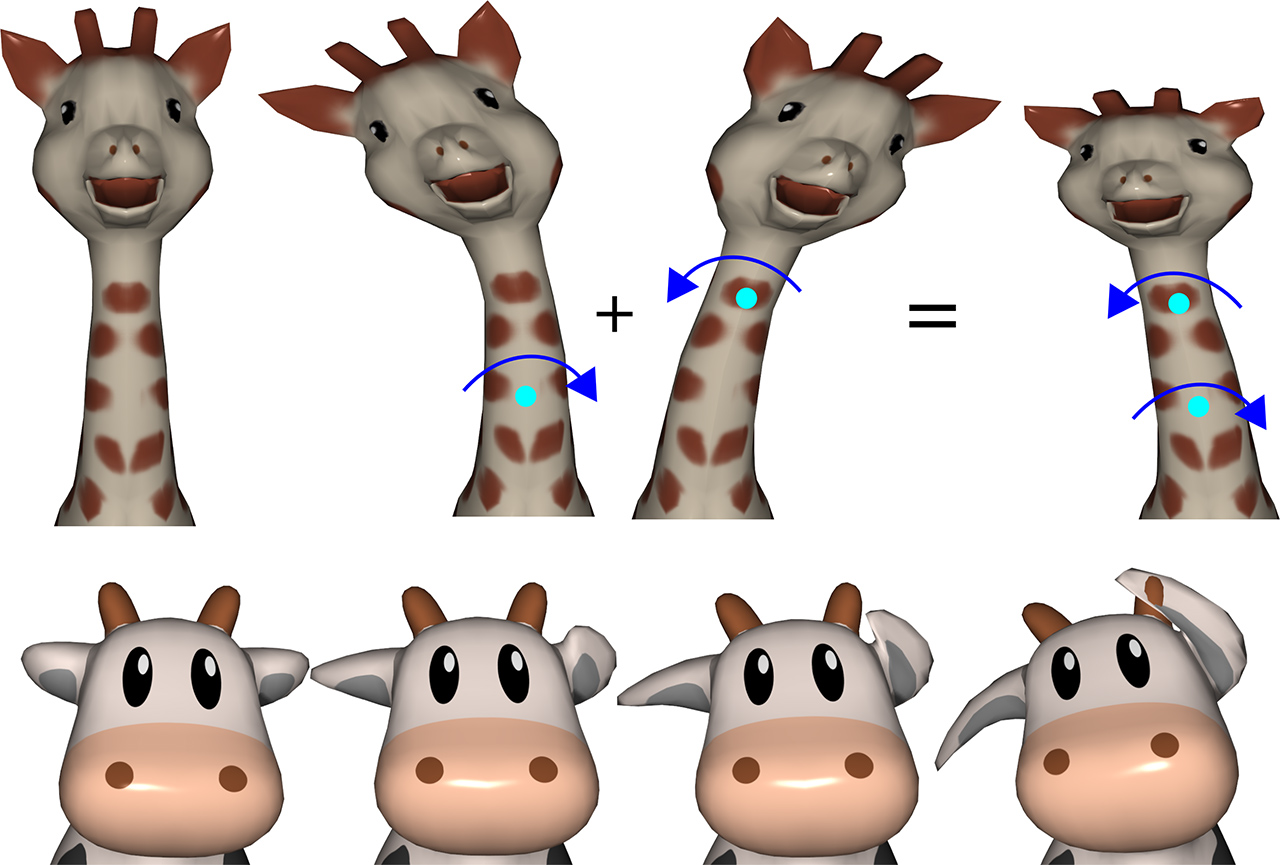}
  \caption{\final{Limitations of the deformers. Top: applying two consecutive and opposite rotations to the giraffe neck induces scaling artifacts on its head. Bottom: The highly floppy ears of the cow self intersect with the geometry when increasing the velocity of motion.}}
  \label{fig:limitation}
  \vspace{-0.5cm}
\end{figure}

To be clear, our method is not physically based and therefore the animations we produce will not necessarily be realistic. Velocity skinning is not intended as a replacement for dynamic simulation, although it may look similar to it when well tuned. 
A \final{benefit} of our method compared to simulation is that any per-vertex displacement is applied instantaneously, and can be computed independently for each frame.
We hypothesize that accounting for acceleration terms in addition to velocity will narrow the gap, and we plan to experiment with this in future work. Another future research direction would be to apply data-driven training to the models, in order to make the velocity skinning weights match a physical simulation.
\final{Visual appearance may degrade under various settings, especially when summing different contributions in 
%opposing 
opposite directions.  For example, applying fast and opposite rotations on successive joints can result in undesirable scaling. In our experiments, such 
%degradations 
artifacts were only visible in carefully crafted and exaggerated 
%design 
examples such as the one 
%illustrated 
shown in Fig~\ref{fig:limitation}. Otherwise, the animation remained surprisingly well-behaved, without visible artifacts, even when the deformation was quite large. 
An additional limitation of the current implementation is that motion applied to the end effectors of a character will not propagate backward within the hierarchy - this effect would require dynamic adaptation of the skeleton hierarchy. Therefore, our method cannot achieve contact-like effects and IK-guided motion (contrasting further with~\cite{zhang2020}).
}

From the \emph{end-application} point of view, 
velocity skinning adds a only limited overhead on top of standard LBS: both are single-pass techniques that fit interactive applications as well as a GPU pipeline. The largest impact stems from the fact that weights \(\AncWeiViBi\) are not as sparse as \(\WeiViBi\), because they are propagated along the skeleton hierarchy. As such, velocity skinning cannot offer the same range of optimizations available for standard skinning, where the number of bone-links per vertex is commonly limited to a constant between 2 and 8. %But such limit may only reveal to be effective over a few million of vertices, which still fit to large range of usage.
 %
 %
%One can also note that while velocity skinning cannot model all dynamic effect and physics-inspired guarantees of iterative approaches such as projective dynamics, it remains less costly, even from a straightforward implementation. 
 %
%the computational and memory burden of velocity skinning is roughly comparable to that of basic skinning: both are single-pass techniques that fit interactive applications as well as a GPU pipeline.
Still, the technique is amenable in practice for real-time resource critical settings, as indicated by our implementations.
Further research could be developed to allow a fast, closed-form update of velocity-skinned normals for accurate illumination, 
%quickly update normal directions in closed form, for a velocity-skinned mesh, 
as can be done with basic skinning \cite{skinNormals}.
%(in our demo, normals are recomputed on the deformed geometry in a second pass). %not the case anymore with GPU

For velocity skinning to be generally adopted, it also needs to be amenable to be included in existing \emph{asset-creation pipelines}. 
As a first exploratory step, we made efforts to offer specialized interface handles for animation control. While we were able to solicit limited animator help in the production of some animations in the corresponding video, and we received very positive feedback, more extensive testing
in the field would help identify bottlenecks and give stronger evidence for the utility and benefits of the approach. While we feel
the introduction of the fundamental approach is a stand-alone contribution and we have made some effort to offer a practical implementation for animation control, extensive user testing is an important next step.

In summary, velocity skinning offers a new path to include automatic, stylized deformations and holds great potential for both offline and interactive future animation effects.

\section*{Acknowledgement}

\final{We acknowledge and thank Paul Kry, Kenny Erleben, and all attendees of the 2020 McGill University Bellairs workshop, the origin of this work, for the fruitful discussions and input. We would also like to thank the artists, Rodney Florencio Da Costa, Kayla Rutherford, and Mohammad Saffar that offered their help and feedback in our usability study.\\
\begin{minipage}{0.15\linewidth}
\includegraphics[width=0.9\textwidth]{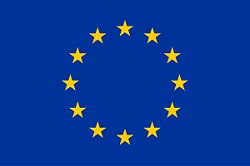}
\end{minipage}
\begin{minipage}{0.85\linewidth}
This project has received funding from the European Union’s Horizon 2020 research and innovation programme under the Marie Sklodowska-Curie grant agreement No. 764644.
\end{minipage}
This paper only contains the author's views and the Research Executive Agency and the Commission are not responsible for any use that may be made of the information it contains.}

%-------------------------------------------------------------------------
% bibtex
\bibliographystyle{eg-alpha-doi} 
\bibliography{references}       

\newcommand{\etalchar}[1]{$^{#1}$}
\begin{thebibliography}{\uppercase{PMRMB15}}

\bibitem[AS07]{angelidis_sca_2007}
\textsc{Angelidis A., Singh K.}:
\newblock {Kinodynamic skinning using volume-preserving deformations}.
\newblock In \emph{SCA} (2007).

\bibitem[BCK{\etalchar{*}}13]{Benard_SIGGRAPH_2013}
\textsc{B\'enard P., Cole F., Kass M., Mordatch I., Hegarty J., Senn M.~S.,
  Fleischer K., Pesare D., Breeden K.}:
\newblock {Stylizing Animation By Example}.
\newblock \emph{ACM Trans. on Graphics. Proc. ACM SIGGRAPH 32}, 4 (2013).

\bibitem[BKLP16]{Bai_SIGGRAPH_2016}
\textsc{Bai Y., Kaufman D.~M., Liu K., Popovic J.}:
\newblock {Artist-Directed Dynamics for 2D Animation}.
\newblock \emph{ACM Trans. on Graphics. Proc. ACM SIGGRAPH 35}, 4 (2016).

\bibitem[BLCD02]{Bregler_2002}
\textsc{Bregler C., Loeb L., Chuang E., Deshpande H.}:
\newblock {Turning to the Masters: Motion Capturing Cartoons}.
\newblock In \emph{SIGGRAPH} (2002).

\bibitem[BML{\etalchar{*}}14]{bouaziz2014}
\textsc{Bouaziz S., Martin S., Liu T., Kavan L., Pauly M.}:
\newblock {Projective dynamics: fusing constraint projections for fast
  simulation}.
\newblock \emph{ACM Trans. on Graphics. Proc. ACM SIGGRAPH 33}, 4 (2014).

\bibitem[CBC{\etalchar{*}}05]{capell_sca_2005}
\textsc{Capell S., Burkhart M., Curless B., Duchamp T., , Popovic Z.}:
\newblock Physically based rigging for deformable characters.
\newblock In \emph{SCA} (2005).

\bibitem[CGC{\etalchar{*}}02]{capell_siggraph_2002}
\textsc{Capell S., Green S., Curless B., Duchamp T., Popovic Z.}:
\newblock {Interactive Skeleton-Driven Dynamic Deformations}.
\newblock \emph{ACM SIGGRAPH} (2002).

\bibitem[CMT{\etalchar{*}}12]{Coros_SIGGRAPH_2012}
\textsc{Coros S., Martin S., Thomaszewski B., Schumacher C., Sumner R., Gross
  M.}:
\newblock {Deformable Objects Alive!}
\newblock \emph{ACM Trans. on Graphics. Proc. ACM SIGGRAPH 31}, 4 (2012).

\bibitem[dASTH10]{aguiar2010}
\textsc{de~Aguiar E., Sigal L., Treuille A., Hodgins J.~K.}:
\newblock {Stable Spaces for Real-time Clothing}.
\newblock \emph{ACM Trans. on Graphics. Proc. ACM SIGGRAPH 29}, 4 (2010).

\bibitem[DB13]{deul_vriphys_2013}
\textsc{Deul C., Bender J.}:
\newblock Physically-based character skinning.
\newblock In \emph{VRIPhys} (2013).

\bibitem[DBB{\etalchar{*}}17]{Dvoroznak_SIGGRAPH_2017}
\textsc{Dvoroznak M., B\'enard P., Barla P., Wang O., Sykora D.}:
\newblock {Example-Based Expressive Animation of 2D Rigid Bodies}.
\newblock \emph{ACM Trans. on Graphics. Proc. ACM SIGGRAPH 36}, 4 (2017).

\bibitem[FOKGM07]{forstmann2007sca}
\textsc{Forstmann S., Ohya J., Krohn-Grimberghe A., McDougall R.}:
\newblock {Deformation Styles for Spline-based Skeletal Animation}.
\newblock In \emph{SCA} (2007).

\bibitem[GDO07]{Garcia_VRIPHYS_2007}
\textsc{Garcia M., Dingliana J., O'Sullivan C.}:
\newblock {A Physically Based Deformation Model for Interactive Cartoon
  Animation}.
\newblock In \emph{VRIPHYS} (2007).

\bibitem[Gri18]{cthulhu}
\textsc{Grippin A.}:
\newblock \emph{Flying Cthulhu}, 2018.
\newblock URL: \url{https://sketchfab.com}.

\bibitem[HTCS13]{hahn_sca_2013}
\textsc{Hahn F., Thomaszewski B., Coros S., Sumner R.~W.}:
\newblock {Efficient Simulation of Secondary Motion in Rig-Space}.
\newblock In \emph{SCA} (2013).

\bibitem[JBK{\etalchar{*}}12]{jacobson2012}
\textsc{Jacobson A., Baran I., Kavan L., Popovic J., Sorkine O.}:
\newblock {Fast Automatic Skinning Transformations}.
\newblock \emph{ACM Trans. on Graphics. Proc. ACM SIGGRAPH 31}, 4 (2012).

\bibitem[JDKL14]{jacobson_star_2014}
\textsc{Jacobson A., Deng Z., Kavan L., Lewis J.}:
\newblock {Skinning: Real-time Shape Deformation}.
\newblock In \emph{ACM SIGGRAPH 2014 Courses} (2014).

\bibitem[JP02]{james_tog_2002}
\textsc{James D.~L., Pai D.~K.}:
\newblock {DyRT: dynamic response textures for real time deformation simulation
  with graphics hardware}.
\newblock \emph{ACM Trans. on Graphics 21}, 3 (2002).

\bibitem[JS11]{jacobson2011stretchable}
\textsc{Jacobson A., Sorkine O.}:
\newblock Stretchable and twistable bones for skeletal shape deformation.
\newblock In \emph{Proceedings of the 2011 SIGGRAPH Asia Conference} (2011),
  pp.~1--8.

\bibitem[KA08]{kass2008}
\textsc{Kass M., Anderson J.}:
\newblock {Animating Oscillatory Motion With Overlap: Wiggly Splines}.
\newblock \emph{ACM Trans. on Graphics. Proc. ACM SIGGRAPH} (2008).

\bibitem[KB18]{komaritzan_i3d_2018}
\textsc{Komaritzan M., Botsh M.}:
\newblock {Projective Skinning}.
\newblock \emph{PACM, Proc. I3D 1}, 1 (2018).

\bibitem[KCGF14]{Kazi_UIST_2014}
\textsc{Kazi R.~H., Chevalier F., Grossman T., Fitzmaurice G.}:
\newblock {Kitty: Sketching Dynamic and Interactive Illustrations}.
\newblock In \emph{User Interface Software \& Technology} (2014).

\bibitem[KCJL06]{Kim_CGI_2006}
\textsc{Kim J.-H., Choi J.-J., Joon H., Lee I.-K.}:
\newblock {Anticipation Effect Generation for Character Animation}.
\newblock \emph{LNCS, Proc. CGI}, 4035 (2006).

\bibitem[KC{\v{Z}}O08]{kavan2008geometric}
\textsc{Kavan L., Collins S., {\v{Z}}{\'a}ra J., O'Sullivan C.}:
\newblock Geometric skinning with approximate dual quaternion blending.
\newblock \emph{ACM Transactions on Graphics (TOG) 27}, 4 (2008), 1--23.

\bibitem[KGUF16]{Kazi_CHI_2016}
\textsc{Kazi R.~H., Grossman T., Umetani B., Fitzmaurice G.}:
\newblock {Motion Amplifiers: Sketching Dynamic Illustrations Using the
  Principles of 2D Animation}.
\newblock \emph{ACM CHI} (2016).

\bibitem[KJP02]{Kry_SCA_2002}
\textsc{Kry P.~G., James D.~L., Pai D.~K.}:
\newblock {EigenSkin: real time large deformation character skinning in
  hardware}.
\newblock In \emph{SCA} (2002).

\bibitem[KL08]{Kwon_CGF_2008}
\textsc{Kwon J., Lee I.}:
\newblock {Exaggerating Character Motions Using Sub-Joint Hierarchy}.
\newblock \emph{Computer Graphics Forum. 27}, 6 (2008), 1677--1686.

\bibitem[KL12]{Yong_TVCG_2012}
\textsc{Kwon J.-Y., Lee I.-K.}:
\newblock {The Squash-and-Stretch Stylization for Character Motions}.
\newblock \emph{IEEE TVCG 18}, 3 (2012).

\bibitem[KS12]{kavan_siggrapha_2012}
\textsc{Kavan L., Sorkine O.}:
\newblock {Elasticity-Inspired Deformers for Character Articulation}.
\newblock \emph{ACM Trans. on Graphics. Proc. ACM SIGGRAPH Asia 31}, 6 (2012).

\bibitem[Las87]{Lasseter}
\textsc{Lasseter J.}:
\newblock {Principles of Traditional Animation Applied to 3D Computer
  Animation}.
\newblock \emph{Computer Graphics. Proc. ACM SIGGRAPH 21}, 4 (1987).

\bibitem[LCa05]{larboulette_2005}
\textsc{Larboulette C., Cani M.-P., arnaldi B.}:
\newblock {Dynamic Skinning: Adding Real-time Dynamic Effects to an Existing
  Character Animation}.
\newblock In \emph{Spring Conference on Computer Graphics} (2005).

\bibitem[LCF00]{Lewis_siggraph_2000}
\textsc{Lewis J., Cordner M., Fong N.}:
\newblock {Pose Space Deformation: A Unified Approach to Shape Interpolation
  andSkeleton-Driven Deformation}.
\newblock \emph{ACM SIGGRAPH} (2000).

\bibitem[LGXS03]{Li_SCA_2003}
\textsc{Li Y., Gleicher M., Xu Y.-Q., Shum H.-Y.}:
\newblock {Stylizing Motion with Drawings}.
\newblock In \emph{SCA} (2003).

\bibitem[LL19]{huy_2019_siggraph_ddm}
\textsc{Le B.~H., Lewis J.}:
\newblock {Direct Delta Mush Skinning and Variants}.
\newblock \emph{ACM Trans. on Graphics. Proc. ACM SIGGRAPH 38}, 4 (2019).

\bibitem[LYKL12]{Young_GM_2012}
\textsc{Lee S.-Y., Yoon J.-C., Kwon J.-Y., Lee I.-K.}:
\newblock {CartoonModes: Cartoon stylization of video objects through modal
  analysis}.
\newblock \emph{Graphical Models 74} (2012), 51--60.

\bibitem[May18]{maya_deformer}
\textsc{Maya}:
\newblock \emph{Squash and Jiggle deformers}.
\newblock Autodesk, 2018.
\newblock https://knowledge.autodesk.com.

\bibitem[MDR{\etalchar{*}}14]{mancewicz_2014_digipro_delta_mush}
\textsc{Mancewicz J., Derksen M.~L., Rijpkema H., , Wilson. C.~A.}:
\newblock {DeltaMush: Smoothing Deformations While Preserving Detail}.
\newblock In \emph{DigiPro} (2014).

\bibitem[MG03]{mohr2003}
\textsc{Mohr A., Gleicher M.}:
\newblock {Building Efficient, Accurate Characer Skins from Examples}.
\newblock \emph{ACM Trans. on Graphics. Proc. ACM SIGGRAPH 22}, 3 (2003).

\bibitem[MHTG05]{muller2005}
\textsc{Muller M., Heidelberger B., Teschner M., Gross M.}:
\newblock {Meshless deformations based on shape matching}.
\newblock \emph{ACM Trans. on Graphics. Proc. ACM SIGGRAPH 24}, 3 (2005).

\bibitem[MK16]{mukai2016}
\textsc{Mukai T., Kuriyama S.}:
\newblock {Efficient Dynamic Skinning with Low-Rank Helper Bone Controllers}.
\newblock \emph{ACM Trans. on Graphics. Proc. ACM SIGGRAPH 35}, 4 (2016).

\bibitem[MMC16]{macklin2016}
\textsc{Macklin M., Muller M., Chentanez N.}:
\newblock {XPBD: Position-Based Simulation of Compliant Constrained Dynamics}.
\newblock \emph{MIG} (2016).

\bibitem[MMG06]{merry2006animation}
\textsc{Merry B., Marais P., Gain J.}:
\newblock Animation space: A truly linear framework for character animation.
\newblock \emph{ACM Transactions on Graphics (TOG) 25}, 4 (2006), 1400--1423.

\bibitem[MTLT88]{magnenat1988joint}
\textsc{Magnenat-Thalmann N., Laperrire R., Thalmann D.}:
\newblock Joint-dependent local deformations for hand animation and object
  grasping.
\newblock In \emph{In Proceedings on Graphics interface’88} (1988), Citeseer.

\bibitem[MZSE11]{mcadams_siggraph_2011}
\textsc{McAdams A., Zhu Y., Selle A., Empey M.}:
\newblock {Efficient elasticityfor character skinning with contact and
  collisions}.
\newblock \emph{ACM Trans. on Graphics. Proc. ACM SIGGRAPH 30}, 4 (2011).

\bibitem[{Nao}15]{iwamoto2015}
\textsc{{Naoya Iwamoto and Hubert P.H. Shum and Longzhi Yang and Shigeo
  Morishima}}:
\newblock {Multi-layer Lattice Model for Real-Time Dynamic Character
  Deformation}.
\newblock \emph{Computer Graphics Forum. Proc. Pacific Graphics 34}, 7 (2015).

\bibitem[NFB16]{nieto_siggraph_talk_2016}
\textsc{Nieto J.~R., Facey T., Brugnot S.}:
\newblock {A Flexible Rigging Framework for VFX and Feature Animation}.
\newblock In \emph{ACM SIGGRAPH Talks} (2016).

\bibitem[NSACO05]{nealen_siggraph_2005}
\textsc{Nealen A., Sorkine O., Alexa M., Cohen-Or D.}:
\newblock {A Sketch-Based Interface for Detail-Preserving Mesh Editing}.
\newblock \emph{ACM Trans. on Graphics. Proc. ACM SIGGRAPH 24}, 3 (2005).

\bibitem[NT06]{Noble_GRAPHITE_2006}
\textsc{Noble P., Tang W.}:
\newblock {Automatic Expressive Deformations for Stylizing Motion}.
\newblock In \emph{GRAPHITE} (2006).

\bibitem[OM94]{Opalach_WorkshopAnimSimu_1994}
\textsc{Opalach A., Maddock S.}:
\newblock {Disney Effects Using Implicit Surfaces}.
\newblock In \emph{Workshop on Animation and Simulation} (1994).

\bibitem[OZ10]{ves_handbook}
\textsc{Okun J.~A., Zwerman S.}:
\newblock \emph{{The VES Handbook of Visual Effects. Industry Standard VFX
  Practices and Procedures}}.
\newblock Focal Press, 2010.

\bibitem[PMRMB15]{pons-moll2015}
\textsc{Pons-Moll G., Romero J., Mahmood N., Black M.~J.}:
\newblock {Dyna: A Model of Dynamic Human Shape in Motion}.
\newblock \emph{ACM Trans. on Graphics. Proc. ACM SIGGRAPH 34}, 4 (2015).

\bibitem[RF14]{Rumman_PPD_skinning_2014}
\textsc{Rumman N.~A., Fratarcangeli M.}:
\newblock Position based skinning of skeleton-driven deformable characters.
\newblock In \emph{Proceedings of the 30th Spring Conference on Computer
  Graphics} (2014), SCCG ’14, p.~83–90.

\bibitem[RHC09]{rohmer_sca_2009}
\textsc{Rohmer D., Hahmann S., Cani. M.-P.}:
\newblock {Exact volume preserving skinning with shape control}.
\newblock In \emph{SCA} (2009).

\bibitem[RL13]{ramos_2013}
\textsc{Ramos J., Larboulette C.}:
\newblock {A Muscle Model for Enhanced Character Skinning}.
\newblock \emph{Journal of WSCG 21}, 2 (2013).

\bibitem[RM13]{Robert_ICIVC_2013}
\textsc{Roberts R., Mallett B.}:
\newblock {A Pose Space for Squash and Stretch Deformation}.
\newblock In \emph{Int. Conf. on Image and Vision Computing} (2013).

\bibitem[RPM]{Ruhland_CS_2017}
\textsc{Ruhland K., Prasad M., McDonnel R.}:
\newblock Data-driven approach to synthesizing facial animation using motion
  capture.

\bibitem[RRC{\etalchar{*}}18]{rousselet_2018}
\textsc{Rousselet V., Rumman N.~A., Canzin F., Mellado N., Kavan L., Barthe
  L.}:
\newblock Dynamic implicit muscles for character skinning.
\newblock \emph{Computer \& Graphics 77} (2018).

\bibitem[TJ81]{illusionOfLife}
\textsc{Thomas F., Johnston O.}:
\newblock \emph{Disney Animation: The illusion of life}.
\newblock Disney Editions, 1981.

\bibitem[TPSH14]{skinNormals}
\textsc{Tarini M., Panozzo D., Sorkine-Hornung O.}:
\newblock Accurate and efficient lighting for skinned models.
\newblock \emph{Comput. Graph. Forum 33}, 2 (2014), 421–428.

\bibitem[VBG{\etalchar{*}}13]{vaillant_2013_siggraph_implicit_skinning}
\textsc{Vaillant R., Barthe L., Guennebaud G., Cani M.-P., Rohmer D., Wyvill
  B., Gourmel O., , Paulin M.}:
\newblock {Implicit Skinning: Real-time Skin Deformation with Contact
  Modeling}.
\newblock \emph{ACM Trans. on Graphics. Proc. ACM SIGGRAPH 32}, 4 (2013).

\bibitem[VGB{\etalchar{*}}14]{vaillant:Siggraph-Asia-2014}
\textsc{Vaillant R., Guennebaud G., Barthe L., Wyvill B., Cani M.-P.}:
\newblock {Robust iso-surface tracking for interactive character skinning}.
\newblock \emph{{ACM Transactions on Graphics} 33}, 6 (Nov. 2014), 1 -- 11.

\bibitem[WDAC06]{Wang_SIGGRAPH_2006}
\textsc{Wang J., Drucker S.~M., Agrawala M., Cohen M.~F.}:
\newblock The cartoon animation filter.
\newblock \emph{ACM Trans. on Graphics. Proc. ACM SIGGRAPH 25} (2006).

\bibitem[Wil01]{AnimatorHandbook}
\textsc{Williams R.}:
\newblock \emph{The animator's survival kit}.
\newblock Faber and Faber, 2001.

\bibitem[WJBK15]{wang2015}
\textsc{Wang Y., Jacobson A., Barbic J., Kavan L.}:
\newblock {Linear Subspace Design for Real-Time Shape Deformation}.
\newblock \emph{ACM Trans. on Graphics. Proc. ACM SIGGRAPH 34}, 4 (2015).

\bibitem[WP02]{wang2002multi}
\textsc{Wang X.~C., Phillips C.}:
\newblock Multi-weight enveloping: least-squares approximation techniques for
  skin animation.
\newblock In \emph{Proceedings of the 2002 ACM SIGGRAPH/Eurographics symposium
  on Computer animation} (2002), pp.~129--138.

\bibitem[WWB{\etalchar{*}}19]{RedMax2019}
\textsc{Wang Y., Weidner N.~J., Baxter M.~A., Hwang Y., Kaufman D.~M., Sueda
  S.}:
\newblock \textsc{RedMax}: Efficient \& flexible approach for articulated
  dynamics.
\newblock \emph{{ACM} Trans.\ Graph. 38}, 4 (July 2019).

\bibitem[XB16]{xu_siggraph_2016}
\textsc{Xu H., Barbic J.}:
\newblock {Pose-Space Subspace Dynamics}.
\newblock \emph{ACM Trans. on Graphics. Proc. ACM SIGGRAPH 35}, 4 (2016).

\bibitem[YSZ06]{yang2006curve}
\textsc{Yang X., Somasekharan A., Zhang J.~J.}:
\newblock {{Curve skeleton skinning for human and creature characters}}.
\newblock \emph{Computer Animation \& Virtual Worlds} (2006).

\bibitem[ZBLJ20]{zhang2020}
\textsc{Zhang J.~E., Bang S., Levin D., Jacobson A.}:
\newblock {Complementary Dynamics}.
\newblock \emph{ACM Trans. on Graphics. Proc. ACM SIGGRAPH Asia} (2020).

\end{thebibliography}

% biblatex with biber
% \printbibliography                

%-------------------------------------------------------------------------

\appendix
\section{Derivation of Equation~(\ref{eqVelSkinning2})}
%from (\ref{eqVelSkinning}) 
\label{sec:derive}
Starting from Equation~\eqref{eqVelSkinning}:
\begin{equation*}
\begin{split}
 \VelVi 
 & = \sum_\Bi{ \WeiViBi  \left( \sum_{\Bj\in \Anc(\Bi)}{ \VelViBj }  \right) } \\
 & = \sum_\Bi{ \sum_{\Bj\in \Anc(\Bi)}{ \WeiViBi \, \VelViBj } } \\
 & = \sum_\Bi{ \sum_{\Bj} { B\left(\,\Bj \in \Anc(\Bi)\,\right) \, \WeiViBi \VelViBj } } \\
 & = \sum_\Bj{ \sum_{\Bi} { B\left(\,\Bj \in \Anc(\Bi)\,\right) \, \WeiViBi \VelViBj } } \\
 & = \sum_\Bj{ \sum_{\Bi} { B\left(\,\Bi \in \Des(\Bj)\,\right) \, \WeiViBi \VelViBj } } \\
 & = \sum_\Bj{ \sum_{\Bi \in \Des(\Bj)} { \WeiViBi \VelViBj } } \\
 & = \sum_\Bj{ \VelViBj \left( \sum_{\Bi \in \Des(\Bj)} {  \WeiViBi  } \right) } 
\end{split}
\end{equation*}
\noindent
Where $B$ is the conditional function, such that $B(\mathrm{true})=1$ and $B(\mathrm{false})=0$, 
and, the 5th line uses the equivalence $$ \Bj \in \Anc(\Bi) \Leftrightarrow \Bi \in \Des(\Bj) $$
At the end, renaming $(i,j)$ as $(j,i)$, and substituting with Equation~\eqref{eq:AncWei}, gives Equation~\eqref{eqVelSkinning2}.
\section{Bone-centroids as barycenters} \label{sec:baricente}
Using the simplifying assumption that the mass of the model is only concentrated at its surface, 
we approximate the barycenter of the portion of the model affected by a given bone $\Bi$ as:
\begin{equation}
\label{eqBarycenters}
 \CentroidBi =  
 \left. \sum_\Vi{ \DesWeiViBi \MasVi}  \PosVi \middle/ \sum_\Vi{ \DesWeiViBi \MasVi } \right. 
\end{equation}
where $\MasVi$ is the area of the Voronoi cell associated to vertex $\Vi$ (that is, one third of the areas of all triangles associated to  $\Vi$). 
and $\DesWeiViBi$ is
the downward propagated vector (see Figure~\ref{fig:propweights}, bottom) given by \(\DesWeiViBi =  \sum_{ \Bj \in \Anc(\Bi)}{ \WeiViBj } \)
%\begin{equation}
%\label{eq:DesWei}
% \DesWeiViBi =  \sum_{ \Bj \in \Anc(\Bi)}{ \WeiViBj } 
%\end{equation}

%, and 
%.

%\section{Construction of matrix $\matrix{R}$ in Equation~(\ref{eq:RotFunSquash})}
%\label{sec:deriveR}
%Let $\vec{r}_x$, $\vec{r}_y$ , $\vec{r}_z$ be the three rows of $\matrix{R}$, and $\vec{m}$ the direction of the medial axis. Then:
%\begin{equation*}
%\begin{split}
% \vec{r}_y & = \vec{m} /  \| \vec{m} \| \\
% \vec{r}_x & = ( \vec{r}_y \times  \AngVelBi ) /  \| \vec{r}_y \times  \AngVelBi \| \\
% \vec{r}_z & =   \vec{r}_x \times \vec{r}_y.
%\end{split}
%\end{equation*}

\end{document}